\documentclass{article}
\usepackage{amssymb}

\usepackage{graphicx}
\usepackage{amsmath}

\input{tcilatex}

\begin{document}

\title{Field quantization, photons and non-Hermitean modes. }
\author{S. A. Brown$^{(a)}$ and B. J. Dalton$^{(a),(b)}$ \\
(a) Department of Physics, University of Queensland, St Lucia, \\
Queensland 4072, Australia\\
(b) Sussex Centre for Optical and Atomic Physics,\\
University of Sussex, Brighton BN1 9QH, United Kingdom}
\maketitle

\begin{abstract}
Field quantization in unstable optical systems is treated by expanding the
vector potential in terms of non-Hermitean (Fox-Li) modes. We define
non-Hermitean modes and their adjoints in both the cavity and external
regions and make use of the important bi-orthogonality relationships that
exist within each mode set. We employ a standard canonical quantization
procedure involving the introduction of generalised coordinates and momenta
for the electromagnetic (EM) field. Three dimensional systems are treated,
making use of the paraxial and monochromaticity approximations for the
cavity non-Hermitean modes. We show that the quantum EM field is equivalent
to a set of quantum harmonic oscillators (QHO), associated with either the
cavity or the external region non-Hermitean modes, and thus confirming the
validity of the photon model in unstable optical systems. Unlike in the
conventional (Hermitean mode) case, the annihilation and creation operators
we define for each QHO are not Hermitean adjoints. It is shown that the
quantum Hamiltonian for the EM field is the sum of non-commuting cavity and
external region contributions, each of which can be expressed as a sum of
independent QHO Hamiltonians for each non-Hermitean mode, except that the
external field Hamiltonian also includes a coupling term responsible for
external non-Hermitean mode photon exchange processes. The non-commutativity
of certain cavity and external region annihilation and creation operators is
associated with cavity energy gain and loss processes, and may be described
in terms of surface integrals involving cavity and external region
non-Hermitean mode functions on the cavity-external region boundary. Using
the essential states approach and the rotating wave approximation, our
results are applied to the spontaneous decay of a two-level atom inside an
unstable cavity. We find that atomic transitions leading to cavity
non-Hermitean mode photon absorption are associated with a different
coupling constant to that for transitions leading to photon emission, a
feature consequent on the use of non-Hermitean mode functions. We show that
under certain conditions the spontaneous decay rate is enhanced by the
Petermann factor.
\end{abstract}

\section{Introduction}

\label{SectIntrod}

The photon model of the quantum electromagnetic (EM) field has recently been
developed further for the areas of cavity QED and quantum optics in
dielectric media. Recent work has demonstrated its validity in microscopic
theories \cite{Huttner92}, where EM field quantization is the same as in
free space, and in macroscopic theories, where the classical optical device
or material medium is treated as a spatially dependent permittivity.
Canonical quantization carried out by expanding the vector potential in
terms of the true (or universe) modes for the system (obtained from a
Helmholtz equation) shows that the EM field is again equivalent to uncoupled
quantum harmonic oscillators (QHO), one for each true mode \cite
{Knoll87,Glauber91,Dalton96,Dalton97}. If idealised or quasi modes (as
obtained from an approximate optical system permittivity) are used instead
of true modes, coupled QHO are obtained, one for each quasi mode \cite
{Glauber91,Dalton99a} and photon exchange processes between quasi modes
occur. Phenomena such as beam splitter effects \cite{Dalton99b}, light
energy loss from cavities \cite{Dalton99c}, reflection and refraction \cite
{Brown01b} can be explained via such processes using quasi mode theory. Both
localised quasi mode (as in cavity or external regions for the standard
cavity QED model \cite{Dalton99c}) and non-localised quasi mode (as in the
beam splitter \cite{Dalton99b}) cases occur. Quasi mode theory has been
extended \cite{Brown01a} to cases \cite{Brown01b} where two sets of quasi
modes are needed.

The photon model obtained by quantization based on normal mode (free space,
true or quasi mode) expansions involves mode functions determined from
Hermitean eigenvalue equations and which satisfy standard orthogonality
conditions (power orthogonality). However, for unstable optical resonators
or gain-guided lasers, the natural EM field modes to use (Fox-Li modes) are
eigenfunctions of a non-Hermitian operator. Biorthogonality conditions apply
for these non-Hermitean modes (NHM) and the related set of Hermitean adjoint
mode functions (HAM). The appropriateness of the photon model \cite
{Siegman95} has been questioned for such systems, where phenomena such as
linewidth enhancement in gain-guided lasers occur. This linewidth
enhancement was interpreted \cite{Petermann79} as excess spontaneous
emission into the laser mode and described by the Petermann factor, and
several semiclassical studies (for example \cite{Siegman89,Hamel89,Rippin96}%
) have successfully accounted for the experimentally observed Peterman
factors using the normalisation integrals for the HAM functions (the
original NHM are normalised to unity). A variety of differing quantum
treatments aimed at explaining the linewidth enhancement have been
published. Some have the objective of trying to account for the success of
the semiclassical theories in terms of a fully quantum treatment of the
field, others present other explanations of the excess noise without using
the NHM functions. In \cite{Deutsch91} the enhanced linewidth is attributed
as a propagation effect arising from longitudinally inhomogeneous noise
gain, rather than an excess of local spontaneous emission, the microscopic
rate of spontaneous emission into a given non-power-orthogonal cavity mode
not being enhanced by the Petermann factor. Reference \cite{Goldberg91}
accounts for the excess noise factor via either spontaneous emission noise
or amplification of vacuum modes leaking into the cavity, depending on the
choice made for operator ordering. This treatment did not involve the use of
NHM and single atom decay was found to be enhanced by the cavity Q factor,
not the Petermann factor. In \cite{Grangier98} a simple few mode model was
presented, and excess noise attributed to loss-induced coupling between the
cavity eigenmodes, with no use of NHM being made. In reference \cite
{Lamprecht99} on the other hand, field quantization based on the NHM and the
HAM is described, and the spontaneous emission rate of an atom in the cavity
found to be enhanced by the Petermann factor. The work in \cite{Bardroff99}
is a laser theory involving a reservoir of excited atoms. Starting from a
true mode description, quasimodes and their adjoints satisfying
non-Hermitean eigenvalue equations are introduced, and results approximating
to those for the semiclassical theories are obtained for the noise
amplification. Finally, in reference \cite{Dutra00} field quantization based
on NHM and the HAM obtained by solving the Helmholtz equation in both the
cavity and external regions with non-Hermitean boundary conditions is
presented. As in \cite{Goldberg91} the single atom decay was found to be
enhanced by the cavity Q factor, not the Petermann factor.

The present paper is a further quantum treatment involving a standard
canonical quantization proceedure based on expanding the vector potential in
the unstable cavity region via the non-Hermitean (Fox-Li) modes and their
Hermitean adjoint modes. It is aimed at trying to account for the success of
the semiclassical theories in terms of a fully quantum treatment of the
field. Our approach is similar to those of \cite{Lamprecht99}, \cite{Dutra00}%
, although here we employ the canonical quantization method and also treat
three dimensional systems, using the paraxial \cite{Lax75} and
monochromaticity approximations \cite{Lamprecht99} for the cavity NHM. As in 
\cite{Lamprecht99}, \cite{Dutra00}, we consider all field modes, rather than
a small number as in the simplified model treated in \cite{Grangier98}. Our
NHM and HAM are also obtained from the properties of the unstable optical
system, rather than from treating atom-field interactions, as in \cite
{Bardroff99}. As in \cite{Dutra00}, the external region is treated and the
field described via further NHM and HAM. Similarly to \cite{Lamprecht99}, 
\cite{Dutra00}, we find that the field is equivalent to a set of QHO's,
associated now with non-Hermitean modes rather than true modes and thus
confirming the validity of the photon model for the case of unstable optical
systems. Similarly to \cite{Lamprecht99}, \cite{Dutra00}, the annihilation,
creation operator pairs for each QHO are not Hermitean adjoints. However,
our results differ significantly in detail from those in \cite{Lamprecht99}, 
\cite{Dutra00}. Our canonical quantization procedure involving both right
and left travelling modes leads to a doubling of the annihilation, creation
operator pairs compared to \cite{Lamprecht99}, \cite{Dutra00}. We show that
the total number of true mode QHO's equals the total number of QHO's
associated with either with the cavity or the external region NHM, one
oscillator for each annihilation, creation operator pair. As in \cite
{Dutra00}, the final form of our quantum Hamiltonian for the EM field is the
sum of non-commuting cavity and external region contributions, but is
simpler as there are no off-diagonal terms. To a good approximation, both
the cavity and external region Hamiltonians can be expressed as the sum of
independent QHO Hamiltonians for each NHM, but the external region
Hamiltonian also includes a coupling term responsible for external NHM
photon exchange processes. The two independent QHO Hamiltonians are
alternative choices for an unperturbed Hamiltonian. Analogous to \cite
{Lamprecht99} for the cavity region Hamiltonian, we obtain left and right
eigenstates for each of these unperturbed Hamiltonians and the energy is
given by the usual QHO result. As in \cite{Dutra00}, certain cavity NHM and
external region NHM annihilation, creation operators do not commute, leading
to cavity energy gain and loss processes, though again the details differ.
We find a simple description of the resultant cavity-external region light
coupling in terms of surface integrals involving products of cavity and
external NHM functions. Atom-field interactions are treated in the electric
dipole and rotating wave approximations. We show that atomic transitions
leading to cavity NHM photon absorption are associated with a different
coupling constant to that for atomic transitions leading to photon emission.
This feature is directly related to the treatment being based on NHM
functions and leads to enhanced emission rates. Using the essential states
approach, we consider spontaneous decay of a two level atom located in the
cavity. The coupling term in the field Hamiltonian is neglected, assuming
that atomic decay into the cavity is much faster than cavity decay. Coupled
equations are obtained for atom-field states amplitudes involving the
excited atom and no photons or the ground state atom with one photon in a
cavity NHM. Markovian decay occurs under certain conditions, the decay rate
being enhanced by the Petermann factor \cite{Petermann79} in special cases
when a single NHM dominates.

The plan of this paper is as follows. In Section \ref{SectQuantization}
quantization of the EM field based on NHM is carried out and approximate
energy eigenstates determined. In Section \ref{SectAtom-Field} atom-field
interactions are considered and a simple treatment of spontaneous emission
from a two level atom into an unstable optical cavity is presented. The
results are summarised in Section \ref{SectConc}. The Appendix contains the
detailed calculation of certain commutation rules.

\section{Theory of canonical quantization via non-Hermitean modes}

\label{SectQuantization}

\subsection{\textit{Non-Hermitean mode functions}}

The case of interest is that of unstable optical cavities, for which a
typical example is shown in figure 1. Using the paraxial approximation \cite
{Lax75}, \cite{Milonni88,Siegmann86}, the cavity NHM functions $\mathbf{U}%
_{n}(\mathbf{R})$ and HAM functions $\mathbf{V}_{n}(\mathbf{R})$ describing
right propagating fields associated with light of angular frequency $\omega
_{n}$ can be written: 
\begin{eqnarray}
\mathbf{U}_{n}(\mathbf{R}) &=&\mathcal{N\,}\mathbf{\hat{\alpha}}_{n}\mathbf{%
.\exp (}ik_{n}z).u_{n}(\mathbf{s,}z)  \label{U.eq} \\
\mathbf{V}_{n}(\mathbf{R}) &=&\mathcal{N\,}\mathbf{\hat{\alpha}}_{n}\mathbf{%
.\exp (}ik_{n}z).v_{n}(\mathbf{s,}z),  \label{V.eq}
\end{eqnarray}
with the polarization vectors chosen as $\mathbf{\hat{\alpha}}_{n}\mathbf{=(%
\hat{x}}$ or $\mathbf{\hat{y})}$ and the wave number is $k_{n}=\omega _{n}/c$
$(\geq 0)$. $\mathcal{N\,}$\ is a normalization factor. The fields $u_{n}(%
\mathbf{s,}z)$ and $v_{n}(\mathbf{s,}z)$ are slowly varying functions of $z$
describing the transverse $\mathbf{s=(}x,y)$ dependence of the NHM
functions. These fields satisfy the right propagating paraxial wave
equation: 
\begin{equation}
(\nabla _{T}^{2}+2ik_{n}\partial /\partial z)\;u_{n}(\mathbf{s,}z)=(\nabla
_{T}^{2}+2ik_{n}\partial /\partial z)\;v_{n}(\mathbf{s,}z)=0,
\label{Parax.eq}
\end{equation}
where $\mathbf{\nabla }_{T}=\mathbf{\hat{x}\;}\partial /\partial x+\mathbf{%
\hat{y}\;}\partial /\partial y$ is the transverse component of $\mathbf{%
\nabla }$. In addition, $u_{n}(\mathbf{s,}z)$ and $v_{n}(\mathbf{s,}z)$
satisfy eigenvalue equations of the form: 
\begin{eqnarray}
\hat{\pounds }\;u_{n}(\mathbf{s,}z) &=&\gamma _{n}\;u_{n}(\mathbf{s,}z)
\label{ueig.eq} \\
\hat{\pounds }^{\dagger }\;v_{n}(\mathbf{s,}z) &=&\gamma _{n}^{\ast }\;v_{n}(%
\mathbf{s,}z),  \label{veig.eq}
\end{eqnarray}
where $\hat{\pounds }$ is a two dimensional non-Hermitean round-trip
propagation operator describing right travelling light, $\hat{\pounds }%
^{\dagger }$ is the related Hermitean adjoint operator related to the
transpose operator $\hat{\pounds }^{T}$ describing left travelling light via 
$\hat{\pounds }^{\dagger }=(\hat{\pounds }^{T})^{\ast }$, and $\gamma _{n}$
are the complex eigenvalues which are independent of $z$. The operators $%
\hat{\pounds }$, $\hat{\pounds }^{T}$ and $\hat{\pounds }^{\dagger }$ are
linear integral operators that transform a field in the $z$ plane into a new
field in the same plane, and are defined in terms of the propagation kernel $%
G_{+}(\mathbf{s,}z;\mathbf{s}_{0}\mathbf{,}z)$ such that for any arbitary
transverse function $f(\mathbf{s,}z)$ we have: 
\begin{eqnarray}
g(\mathbf{s,}z) &=&\hat{\pounds }\;f(\mathbf{s,}z)=\int d^{2}\mathbf{s}%
_{0}\,G_{+}(\mathbf{s,}z;\mathbf{s}_{0}\mathbf{,}z)\,f(\mathbf{s}_{0}\mathbf{%
,}z)  \label{Loper.eq} \\
h(\mathbf{s,}z) &=&\hat{\pounds }^{T}\;f(\mathbf{s,}z)=\int d^{2}\mathbf{s}%
_{0}\,G_{+}(\mathbf{s}_{0}\mathbf{,}z;\mathbf{s,}z)\,f(\mathbf{s}_{0}\mathbf{%
,}z)  \label{Ltransoper.eq} \\
i(\mathbf{s,}z) &=&\hat{\pounds }^{\dagger }\;f(\mathbf{s,}z)=\int d^{2}%
\mathbf{s}_{0}\,G_{+}^{\ast }(\mathbf{s}_{0}\mathbf{,}z;\mathbf{s,}z)\,f(%
\mathbf{s}_{0}\mathbf{,}z).  \label{Ladjointoper.eq}
\end{eqnarray}
Thus $\hat{\pounds }$ maps $f(\mathbf{s,}z)$ onto $g(\mathbf{s,}z)$, $\hat{%
\pounds }^{T}$ maps $f(\mathbf{s,}z)$ onto $h(\mathbf{s,}z)$ and $\hat{%
\pounds }^{\dagger }$ maps $f(\mathbf{s,}z)$ onto $i(\mathbf{s,}z)$. The
propagation kernel $G_{+}(\mathbf{s,}z;\mathbf{s}_{0}\mathbf{,}z)$ is
obtained from the Fresnel approximation to the Huygens-Fresnel principle,
details are given in references \cite{Milonni88,Siegmann86}. For the
situation depicted in figure 1, $G_{+}$ would be constructed allowing for
right travelling propagation from the $z$ plane to the right end mirror,
left travelling propagation to the left end mirror and then right travelling
propagation to the $z$ plane. The fields $\hat{\pounds }\;f(\mathbf{s,}z)$
and $\hat{\pounds }^{\dagger }\;f(\mathbf{s,}z)$ also satisfy the same
paraxial wave equations as $u_{n}(\mathbf{s,}z)$ and $v_{n}(\mathbf{s,}z)$.
In our notation $n$ will specify $\mathbf{\hat{\alpha}}_{n},k_{n}$ (the
polarization and longitudinal wave number) and the transverse mode index $%
\theta _{n}$ for the distinct transverse NHM functions in which the
quantities $\mathbf{\hat{\alpha}}_{n},k_{n}$ are the same. It will be
convenient to quantize the wave numbers, with $k_{n}=N_{n}\,\pi /\,l$ ,
where $N_{n}$ is an integer and $l$ is the cavity length. Both $\mathbf{U}%
_{n},\mathbf{V}_{n}$ describe right propagating light fields, whereas 
\begin{equation}
\mathbf{W}_{n}=\mathbf{\hat{\alpha}}_{n}\mathbf{.\exp (}ik_{n}z).w_{n}(%
\mathbf{s,}z)=(\mathbf{V}_{n})^{\ast },  \label{W.eq}
\end{equation}
describes left propagating light fields, where $w_{n}(\mathbf{s,}z)=v_{n}(%
\mathbf{s,}z)^{\ast }$ and satisfies the eigenvalue equation $\hat{\pounds }%
^{T}\;w_{n}(\mathbf{s,}z)=\gamma _{n}\;w_{n}(\mathbf{s,}z)$, . Thus we see
that $u_{n}(\mathbf{s,}z)$ and $v_{n}(\mathbf{s,}z)$ are special solutions
of the right propagating paraxial wave equation that are self-reproducing
under the propagation operators $\hat{\pounds }$ and $\hat{\pounds }%
^{\dagger }$ respectively.

Based on the well known two dimensional orthogonality results \cite
{Siegman89} for the $u_{n}$ and $v_{n}$ such as $\int d^{2}\mathbf{s\;}u_{n}(%
\mathbf{s,}z)^{\ast }v_{m}(\mathbf{s,}z)=\delta _{nm}$, we can obtain
important three dimensional orthogonality integrals in the form:

\begin{eqnarray}
\int_{C}d^{3}\mathbf{R\;U}_{n}(\mathbf{R})^{{\Large \ast }}\bullet \mathbf{V}%
_{m}(\mathbf{R}) &=&\delta _{nm}  \label{Biorth.eq} \\
\int_{C}d^{3}\mathbf{R\;U}_{n}(\mathbf{R})^{{\Large \ast }}\bullet \mathbf{U}%
_{m}(\mathbf{R}) &=&\mathbf{C}_{nm}  \label{C.eq} \\
\int_{C}d^{3}\mathbf{R\;V}_{n}(\mathbf{R})^{{\Large \ast }}\bullet \mathbf{V}%
_{m}(\mathbf{R}) &=&\mathbf{D}_{nm},  \label{D.eq}
\end{eqnarray}
the first being the biorthogonality condition, the others giving the overlap
integrals for the NHM and the HAM in terms of the Hermitean, positive
definite transformation matrices $\mathbf{C}$ and $\mathbf{D}$. The
integrals are over the cavity region. It can be shown that $\mathbf{C}_{nm}$
and $\mathbf{D}_{nm}$ are zero unless the angular frequencies $\omega
_{n},\omega _{m}$ and the polarizations $\mathbf{\hat{\alpha}}_{n},\mathbf{%
\hat{\alpha}}_{m}$ are equal, and $\mathbf{C\,D=D\,C=E}$. By convention we
choose $\mathbf{C}_{nn}=1$, and then from Schwarz' inequality it follows
that $\mathbf{D}_{nn}\geqslant 1$. The Petermann factor $K_{n}$ is given by $%
\mathbf{D}_{nn}$ in the semiclassical theories (see \cite
{Siegman89,Hamel89,Rippin96}). The NHM functions satisfy a completeness
relation:

\begin{equation}
\sum_{n}(U_{n}^{\alpha }(\mathbf{R})V_{n}^{\beta }(\mathbf{R}^{\prime })^{%
{\Large \ast }}+V_{n}^{\alpha }(\mathbf{R})^{{\Large \ast }}U_{n}^{\beta }(%
\mathbf{R}^{\prime }))=\delta _{\alpha \beta }\,\delta (\mathbf{R-R}^{\prime
})  \label{Comp.eq}
\end{equation}
for $\mathbf{R,R}^{\prime }$ inside the cavity, and where $U_{n}^{\alpha }(%
\mathbf{R)},V_{n}^{\alpha }(\mathbf{R)}$ are the $\alpha $ components ($%
\alpha ,\beta =x,y$) of the vector fields $\mathbf{U}_{n}(\mathbf{R)},%
\mathbf{V}_{n}(\mathbf{R)}$. Integrals similar to those in equations (\ref
{Biorth.eq}, \ref{C.eq}, \ref{D.eq}) but without the complex conjugation are
ignored due to the product of the two fast-varying $\mathbf{\exp (}ik_{n}z)$
factors approximately averaging to zero. The corresponding completeness
relationship in reference \cite{Lamprecht99} does not include the second
term. There are also interrelationships between the NHM and HAM functions in
terms of the transformation matrices:

\begin{equation}
\mathbf{U}_{n}(\mathbf{R})=\sum_{m}\mathbf{C}_{mn}\,\mathbf{V}_{m}(\mathbf{R}%
)\quad \quad \mathbf{V}_{n}(\mathbf{R})=\sum_{m}\mathbf{D}_{mn}\,\mathbf{U}%
_{m}(\mathbf{R}).  \label{UV.eq}
\end{equation}

\subsection{\textit{Vector potential and generalised coordinates}}

The NHM functions and true mode functions represent monochromatic fields.
For more general time dependent fields we can write the vector potential $%
\mathbf{A(R)}$ as the sum of a right travelling light field $\mathbf{A}_{R}%
\mathbf{(R)}$ and a left travelling field $\mathbf{A}_{R}\mathbf{(R)}^{\ast
} $for $\mathbf{R}$ inside or outside the cavity: 
\begin{equation}
\mathbf{A(R)=A}_{R}\mathbf{(R)+A}_{R}\mathbf{(R)}^{\ast },  \label{A.eq}
\end{equation}
and then expand $\mathbf{A}_{R}\mathbf{(R)}$ as a linear combination of
suitable monochromatic fields. One choice is to expand $\mathbf{A}_{R}$ in
terms of true mode functions $\mathbf{A}_{k}(\mathbf{R)}$ in the right half
space $R$. Corresponding true modes in the left half space are denoted $%
\mathbf{A}_{k\ast }(\mathbf{R)=(A}_{k}(\mathbf{R))}^{\ast }$. These modes
all have the same angular frequency $\omega _{k}$. Alternatively, a similar
expansion can be made (but only \textit{within} the cavity) in terms of
either the NHM functions $\mathbf{U}_{n}(\mathbf{R})$ or the HAM functions $%
\mathbf{V}_{n}(\mathbf{R}).$ The expansion coefficients ($q_{k}$ or $Q_{n}$
or $R_{n}$ for the three choices) act as generalised coordinates specifying
the vector potential in the appropriate region. In the left half space $%
q_{k\ast }=(q_{k})^{\ast }$. Thus:

\begin{eqnarray}
\mathbf{A}_{R}\mathbf{(R)} &=&\sum_{k}^{{R}}q_{k}\,\mathbf{A}_{k}(\mathbf{R)}
\label{ArTrue.eq} \\
\mathbf{A}_{R}\mathbf{(R)} &=&\sum_{n}Q_{n}\,\mathbf{U}_{n}(\mathbf{R)=}%
\sum_{n}R_{n}\,\mathbf{V}_{n}(\mathbf{R),}  \label{ArNHM.eq}
\end{eqnarray}
where $\mathbf{R}$ lies within the cavity in equations (\ref{ArNHM.eq}). The 
$Q_{n},R_{n}$ depend on the $q_{k}$, as the latter specify the field
everywhere via equation (\ref{ArTrue.eq}). We have:

\begin{equation}
Q_{n}=\sum_{k}^{R}\mathbf{\Gamma }_{nk}\,q_{k}\qquad R_{n}=\sum_{k}^{R}%
\mathbf{\Lambda }_{nk}\,q_{k},  \label{QRq.eq}
\end{equation}
where 
\begin{eqnarray}
\mathbf{\Gamma }_{nk} &=&\int_{C}d^{3}\mathbf{R\,V}_{n}(\mathbf{R})^{\ast
}\cdot \mathbf{A}_{k}(\mathbf{R)}  \label{Gamma.eq} \\
\mathbf{\Lambda }_{nk} &=&\int_{C}d^{3}\mathbf{R\,U}_{n}(\mathbf{R})^{\ast
}\cdot \mathbf{A}_{k}(\mathbf{R)}.  \label{Lambda.eq}
\end{eqnarray}
The matrices $\mathbf{\Gamma ,\Lambda }$ have properties $\mathbf{\Gamma }%
^{\dagger }\mathbf{\Lambda }=\mathbf{\Lambda }^{\dagger }\mathbf{\Gamma }=%
\mathbf{E}$ and $(\mathbf{\Lambda }^{\dagger }\mathbf{\Lambda )}_{nm}=%
\mathbf{C}_{nm}$, $\ \ \ \ \ \ (\mathbf{\Gamma }^{\dagger }\mathbf{\Gamma )}%
_{nm}=\mathbf{D}_{nm}$. Interrelationships between the $Q_{n}$ and the $%
R_{n} $ can easily be found in terms of $\mathbf{C,D}$. These are:

\begin{equation}
Q_{n}=\sum_{m}\mathbf{D}_{nm}\,R_{m}\qquad R_{n}=\sum_{m}\mathbf{C}%
_{nm}\,Q_{m}  \label{RQ.eq}
\end{equation}

Being determined from solutions of the paraxial wave equation, the cavity
NHM and HAM functions $\mathbf{U}_{n}$ and $\mathbf{V}_{n}$ actually
represent the lowest order terms in a more general treatment starting from
paraxial fields. Following Ref. \cite{Lax75}, the spatial part of the vector
potential in the Coulomb gauge 
\begin{equation}
\mathbf{\nabla \cdot A}=0  \label{Coulgauge.eq}
\end{equation}
for harmonic solutions of the wave equation (and which represent right
travelling waves), can be expressed as 
\begin{equation}
\mathbf{A}_{R}^{(k_{n})}(\mathbf{R})=\mathbf{\exp (}ik_{n}z).[\mathbf{F}_{T}(%
\mathbf{R})+\mathbf{\hat{z}\,}F_{Z}(\mathbf{R})],  \label{GeneralA.eq}
\end{equation}
where $\mathbf{F}_{T}$ and $\mathbf{\hat{z}\,}F_{Z}$ are transverse and
longitudinal components. $\mathbf{F}_{T}$ and $F_{Z}$ can be expressed as
power series in the small quantity $f=w_{0}/l_{k}$, where $w_{0}$ is the
lateral size for the beam and $l_{k}=k\,(w_{0})^{2}$ is the Fraunhofer
diffraction length. $\mathbf{F}_{T},F_{Z}$ involve even, odd powers of $f$
respectively. The zeroth order term in $\mathbf{F}_{T}$ satisfies the
paraxial wave equation \cite{Lax75}, thereby clarifying the status of this
equation as determining the lowest order contribution to $\mathbf{A}%
_{R}^{(k_{n})}$. The cavity NHM and HAM functions $\mathbf{U}_{n}$ and $%
\mathbf{V}_{n}$ are particular zeroth order solutions for $\mathbf{F}_{T}$.
However, to satisfy the Coulomb gauge condition even in zeroth order, the
first order term from $F_{Z}$ would need to be included. This would replace $%
\mathbf{U}_{n}$ and $\mathbf{V}_{n}$ by: 
\begin{eqnarray}
\mathbf{U}_{n}^{(1)} &=&(\mathbf{U}_{n}+i/k_{n}\,\mathbf{\hat{z}\,}\{\mathbf{%
\nabla }_{T}\cdot \mathbf{U}_{n}\})  \label{Un1.eq} \\
\mathbf{V}_{n}^{(1)} &=&(\mathbf{V}_{n}+i/k_{n}\,\mathbf{\hat{z}\,}\{\mathbf{%
\nabla }_{T}\cdot \mathbf{V}_{n}\}),  \label{Vn1.eq}
\end{eqnarray}
resulting in the gauge condition being satisfied correct to the second order
in $f$. In a higher order treatment we would replace $\mathbf{U}_{n}$ and $%
\mathbf{V}_{n}$ by these expressions, which now would have a small first
order correction in the $\mathbf{\hat{z}}$ direction. However, it is
apparent that the small correction terms are perpendicular to the zeroth
order terms $\mathbf{U}_{n}$ or $\mathbf{V}_{n}$. This results in
expressions to terms in the Lagrangian and Hamiltonian based on NHM
expansions only changing in second order, and such corrections can be
neglected in our results. Although the Coulomb gauge condition plays a
central role in the field quantization presented here, the correction terms%
\textit{\ }in the NHM functions need not concern us further, since our
formal canonical quantization procedure is based on the true mode functions 
\cite{Dalton96}. These are constructed to satisfy the Coulomb gauge
condition for the free space situation that applies both inside and outside
the cavity.

Expansions of the right travelling field $\mathbf{A}_{R}\mathbf{(R)}$ may
also be made in the external region in terms of NHM $\mathbf{U}_{K}(\mathbf{R%
})$ and their HAM $\mathbf{V}_{K}(\mathbf{R})$ as determined from other
non-Hermitean operators, and the expansion coefficients $Q_{K},R_{K}$ act as
generalised coordinates for the field in the external region. We do not need
to specify the actual method for determining these mode functions other than
noting that the paraxial approximation would not apply. Generalisations to
three dimensions of the approach in \cite{Dutra00} might be applied.
However, the external NHM functions should satisfy the Coulomb gauge
condition and the biorthogonality conditions, and be chosen so that the
magnetic energy contribution to the Lagrangian is in diagonal form (see
equation (\ref{Omega2.eq})). Orthogonality, completeness and
interrelationships analogous to equations (\ref{Biorth.eq}, \ref{C.eq}, \ref
{D.eq}, \ref{Comp.eq}, \ref{UV.eq}) occur for the $\mathbf{U}_{K}(\mathbf{R}%
) $ and $\mathbf{V}_{K}(\mathbf{R})$. Equations analogous to equations (\ref
{ArNHM.eq}, \ref{QRq.eq}, \ref{RQ.eq}) apply for the $Q_{K},R_{K}$. The
matrices $\mathbf{C,D,\Gamma ,\Lambda }$ are replaced by $\mathbf{G,H,\Delta
,\Phi }$, where: 
\begin{eqnarray}
\int_{E}d^{3}\mathbf{R\,U}_{K}(\mathbf{R})^{\ast }\cdot \mathbf{V}_{L}(%
\mathbf{R)} &=&\mathbf{\,}\delta _{KL}  \label{Biorth2.eq} \\
\int_{E}d^{3}\mathbf{R\,U}_{K}(\mathbf{R})^{\ast }\cdot \mathbf{U}_{L}(%
\mathbf{R)} &=&\mathbf{G}_{KL}  \label{G2.eq} \\
\int_{E}d^{3}\mathbf{R\,V}_{K}(\mathbf{R})^{\ast }\cdot \mathbf{V}_{L}(%
\mathbf{R)} &=&\mathbf{H}_{KL}  \label{H2.eq}
\end{eqnarray}
and 
\begin{eqnarray}
\mathbf{\Delta }_{Kk} &=&\int_{E}d^{3}\mathbf{R\,V}_{K}(\mathbf{R})^{\ast
}\cdot \mathbf{A}_{k}(\mathbf{R)}  \label{Delta.eq} \\
\mathbf{\Phi }_{Kk} &=&\int_{E}d^{3}\mathbf{R\,U}_{K}(\mathbf{R})^{\ast
}\cdot \mathbf{A}_{k}(\mathbf{R)}.  \label{Phi.eq}
\end{eqnarray}
The integrals are over the external region. The matrices $G,H,\Delta ,\Phi $%
\ satisfy relations analogous to those for $C,D,\Gamma ,\Lambda $\textit{.}

The vector potential in general satifies the wave equation. Since the $Q_{n}$
or $R_{n}$ determine the vector potential inside the cavity, they also
determine $\mathbf{A}$ and its derivatives on the boundary with the external
region. Hence they also determine the field in the external region, which is
specified by the $Q_{K}$ or $R_{K}$. However, the field in both regions is
also specified by the $q_{k}$. Hence the sets $Q_{n}$ or $R_{n}$, $Q_{K}$ or 
$R_{K}$ are equivalent to the $q_{k}$ as generalised coordinates specifying
the vector potential, and these coordinates are interconvertable. The
inverse relationships are:

\begin{equation}
q_{k}=\sum_{l,n}\mathbf{M}_{kl}^{-1}\mathbf{\Lambda }_{nl}^{{\ast }%
}\,Q_{n}=\sum_{l,K}\mathbf{N}_{kl}^{-1}\mathbf{\Phi }_{Kl}^{{\ast }%
}\,Q_{K}=\sum_{l,n}\mathbf{M}_{kl}^{-1}\mathbf{\Gamma }_{nl}^{{\ast }%
}\,R_{n}=\sum_{l,K}\mathbf{N}_{kl}^{-1}\mathbf{\Delta }_{Kl}^{{\ast }}\,R_{K}
\label{qQR.eq}
\end{equation}
where the matrices $\mathbf{M,N}$ are defined by cavity and external region
integrals: 
\begin{eqnarray}
\mathbf{M}_{kl} &=&\int_{C}d^{3}\mathbf{R\,A}_{k}(\mathbf{R})^{\ast }\cdot 
\mathbf{A}_{l}(\mathbf{R)}  \label{M.eq} \\
\mathbf{N}_{kl} &=&\int_{E}d^{3}\mathbf{R\,A}_{k}(\mathbf{R})^{\ast }\cdot 
\mathbf{A}_{l}(\mathbf{R)}.  \label{N.eq}
\end{eqnarray}
Clearly $\mathbf{M}_{kl}+\mathbf{N}_{kl}=\delta _{kl}.$The relationship in
equations (\ref{qQR.eq}) illustrates the point that the number of true modes 
$k$ in the \textit{half }space must equal the number of cavity NHM
designated $n$ \textit{or} the number of external NHM designated $K$. Hence
the total number of true modes in the full space would be equal to \textit{%
twice} the number of cavity or external NHM. Equivalently, the total number
of true modes $\mathbf{A}_{k}$ in the full space would be equal to the
number of cavity right travelling NHM $\mathbf{U}_{n}$ plus left travelling
NHM $\mathbf{W}_{n}=\mathbf{V}_{n}^{\ast }$, or the number of right
travelling $\mathbf{U}_{K}$ and left travelling $\mathbf{V}_{K}^{\ast }$
external NHM. This result is important in comparing the total number of
quantum harmonic oscillators for the field described via true modes with
that when it is described by NHM.

\subsection{\textit{Lagrangian}}

The Lagrangian is defined by: 
\begin{equation}
L=\frac{1}{2}\epsilon _{0}\int d^{3}\mathbf{R\,(\dot{A}(R)}^{2}-c^{2}[%
\mathbf{\nabla \times A(R)]}^{2}).  \label{Lagr.def}
\end{equation}
The true mode expansion gives $L$ in terms of the generalised coordinates $%
q_{k}$ and velocities $\dot{q}_{k}$ (and their complex congugates) in the
form: 
\begin{equation}
L=\epsilon _{0}\sum_{k}^{R}\,(\dot{q}_{k}\dot{q}_{k}^{\ast }-\omega
_{k}^{2}q_{k}q_{k}^{\ast }).  \label{LagrTrue.eq}
\end{equation}
The Lagrangian can also be written as: 
\begin{equation}
L=L_{C}+L_{E},  \label{LagrSum.eq}
\end{equation}
the sum of contributions $L_{C},L_{E}$ from the cavity and external regions.
Using the NHM expansion the cavity contribution $L_{C}$ can be obtained in
terms of either the generalised coordinates $Q_{n}$ and velocities $\dot{Q}%
_{n}$ or the alternative quantities $R_{n}$ and $\dot{R}_{n}$ (and their
complex congugates). The external contribution $L_{E}$ involves $Q_{K},$ $%
\dot{Q}_{K}$ or $R_{K}$, $\dot{R}_{K}$. The expressions for these
Lagrangians are:

\begin{eqnarray}
L_{C} &=&\epsilon _{0}\sum_{n}\,(\dot{Q}_{n}\dot{R}_{n}^{\ast }-\omega
_{n}^{2}Q_{n}R_{n}^{\ast })=\epsilon _{0}\sum_{n}\,(\dot{R}_{n}\dot{Q}%
_{n}^{\ast }-\omega _{n}^{2}R_{n}Q_{n}^{\ast })  \label{Lc.eq} \\
L_{E} &=&\epsilon _{0}\sum_{K}\,(\dot{Q}_{K}\dot{R}_{K}^{\ast }-\omega
_{K}^{2}Q_{K}R_{K}^{\ast })=\epsilon _{0}\sum_{K}\,(\dot{R}_{K}\dot{Q}%
_{K}^{\ast }-\omega _{K}^{2}R_{K}Q_{K}^{\ast }).  \label{Le.eq}
\end{eqnarray}
For the cavity term the magnetic field integral has been evaluated
approximately with derivatives of the slowly varying factors being
neglected. The external NHM and HAM have been chosen so that the matrix $%
\mathbf{\Omega }$ determining the magnetic energy contribution, which is
given by: 
\begin{eqnarray}
\mathbf{\Omega }_{KL} &=&c^{2}\int_{E}d^{3}\mathbf{R\,}[\mathbf{\nabla
\times U}_{K}\mathbf{(R)]}^{\ast }\mathbf{\cdot }[\mathbf{\nabla \times V}%
_{L}\mathbf{(R)]}  \label{Omega.eq} \\
&=&\omega _{K}^{2}\,\delta _{KL}\mathbf{,}  \label{Omega2.eq}
\end{eqnarray}
is in diagonal form with eigenvalues $\omega _{K}^{2}$. For the external NHM
a wave number $k_{K}$ can be defined via $\omega _{K}=c\,k_{K}$.

\subsection{\textit{Generalised momenta}}

The true mode generalised momenta are defined in terms of the Lagrangian and
related to the generalised velocities as: 
\begin{eqnarray}
p_{k} &=&\partial L/\partial \dot{q}_{k}^{\ast }  \label{pk.def} \\
p_{k} &=&\epsilon _{0}\dot{q}_{k}.  \label{pkqkdot.eq}
\end{eqnarray}
In the left half space $p_{k\ast }=(p_{k})^{\ast }$. We can also write $%
p_{k} $ as the sum of cavity and external terms as: 
\begin{eqnarray}
p_{k} &=&p_{k}^{C}+p_{k}^{E}  \label{pkSum.eq} \\
p_{k}^{C} &=&\partial L_{C}/\partial \dot{q}_{k}^{\ast }\qquad
p_{k}^{E}=\partial L_{E}/\partial \dot{q}_{k}^{\ast }  \label{pkCavExt.eq}
\end{eqnarray}
For the cavity NHM it is convenient to introduce generalised momenta defined
in terms of the cavity term in the Lagrangian via: 
\begin{equation}
P_{n}=\partial L_{C}/\partial (\dot{Q}_{n})^{\ast }\qquad S_{n}=\partial
L_{C}/\partial (\dot{R}_{n})^{\ast }  \label{PnSn.eq}
\end{equation}
rather than expressions involving derivatives of the total Lagrangian $L$.
We are able to make this somewhat arbitary choice for $P_{n}$ and $S_{n}$
because our formal quantization process is based on the true mode quantities 
$p_{k}$ and $q_{k}$, and so only the $p_{k}$ need to be defined in the
canonical way. The generalised momenta $P_{n}$ and $S_{n}$ are given by: 
\begin{eqnarray}
P_{n} &=&\epsilon _{0}\sum_{m}\mathbf{C}_{nm}\,\dot{Q}_{m}=\epsilon _{0}\dot{%
R}_{n}  \label{PnQmdotRndot.eq} \\
S_{n} &=&\epsilon _{0}\sum_{m}\mathbf{D}_{nm}\,\dot{R}_{m}=\epsilon _{0}\dot{%
Q}_{n}.  \label{SnRmdotQndot.eq}
\end{eqnarray}
We note that the generalised momentum $P_{n}$ associated with the
generalised coordinate $Q_{n}$ is given in terms of the generalised velocity 
$\dot{R}_{n}$ associated with the alternative coordinate $R_{n}$. The
reverse applies as well, and these features are a direct consequence of the
two mode functions $\mathbf{U}_{n}(\mathbf{R)}$ and $\mathbf{V}_{n}(\mathbf{%
R)}$ not being the same for a non-Hermitean mapping operator $\hat{\pounds }$
, since $u_{n}\neq v_{n}$. For the external region similar considerations
give for the generalised momenta $P_{K}=\epsilon _{0}\dot{R}_{K}$ and $%
S_{K}=\epsilon _{0}\dot{Q}_{K}$. Together the sets of NHM generalised
coordinates and momenta $Q_{n},P_{n}$ or $R_{n},S_{n}$ or $Q_{K},P_{K}$ or $%
R_{K},S_{K}$ are equivalent to the true mode generalised coordinates and
momenta $q_{k},p_{k}$. The cavity $P_{n},S_{n}$ depend on the $p_{k}$. We
have:

\begin{equation}
P_{n}=\sum_{k}^{R}\mathbf{\Lambda }_{nk}\,p_{k}\qquad S_{n}=\sum_{k}^{R}%
\mathbf{\Gamma }_{nk}\,p_{k}  \label{PiTp.eq}
\end{equation}
Interrelationships between the $P_{n}$ and the $S_{n}$ can easily be found
in terms of $\mathbf{C,D}$. These are:

\begin{equation}
P_{n}=\sum_{m}\mathbf{C}_{nm}\,S_{m}\qquad S_{n}=\sum_{m}\mathbf{D}%
_{nm}\,P_{m}.  \label{PiT.eq}
\end{equation}
Equations analogous to equations (\ref{PiTp.eq}, \ref{PiT.eq}) also apply
for the $P_{K},S_{K}$. The overall inverse relationships are:

\begin{equation}
p_{k}=\sum_{l,n}\mathbf{M}_{kl}^{-1}\mathbf{\Gamma }_{nl}^{{\large \ast }%
}\,P_{n}=\sum_{l,K}\mathbf{N}_{kl}^{-1}\mathbf{\Delta }_{Kl}^{{\large \ast }%
}\,P_{K}=\sum_{l,n}\mathbf{M}_{kl}^{-1}\mathbf{\Lambda }_{nl}^{{\large \ast }%
}\,S_{n}=\sum_{l,K}\mathbf{N}_{kl}^{-1}\mathbf{\Phi }_{Kl}^{{\large \ast }%
}\,S_{K}  \label{pPiTau.eq}
\end{equation}

\subsection{\textit{Conjugate momentum field}}

The conjugate momentum field $\mathbf{\Pi (R)}$ is defined as the derivative
of the Lagrangian density with respect to $\mathbf{\dot{A}}$ and is given
by: 
\begin{equation}
\mathbf{\Pi (R)=\varepsilon }_{0}\mathbf{\dot{A}(R)}.  \label{ConjMtField.eq}
\end{equation}
For all positions $\mathbf{R}$ it can be written as the sum of a right
travelling light field $\mathbf{\Pi }_{R}\mathbf{(R)}$ and a left travelling
field $\mathbf{\Pi }_{R}\mathbf{(R)}^{\ast }$. The generalised momenta acts
as expansion coefficients for expressing the conjugate momentum field in
terms of the mode functions. In terms of the true mode and NHM expansions
involving the generalised momenta we find that: 
\begin{eqnarray}
\mathbf{\Pi (R)} &=&\mathbf{\Pi }_{R}\mathbf{(R)+\Pi }_{R}\mathbf{(R)}^{\ast
}  \label{CongMtFieldSum.eq} \\
\mathbf{\Pi }_{R}\mathbf{(R)} &=&\sum_{k}^{{\large R}}p_{k}\,\mathbf{A}_{k}(%
\mathbf{R)}  \label{PiRTrue.eq} \\
\mathbf{\Pi }_{R}\mathbf{(R)} &=&\sum_{n}S_{n}\,\mathbf{U}_{n}(\mathbf{R)=}%
\sum_{n}P_{n}\,\mathbf{V}_{n}(\mathbf{R)}  \label{PiRNHM.eq}
\end{eqnarray}
with $\mathbf{R}$ inside the cavity in equations (\ref{PiRNHM.eq}).
Analogous expressions apply for the external region. Comparing the similar
expression for the vector potential (equations (\ref{ArNHM.eq})) we note the
unexpected feature that the momentum $S_{n}$ is now the expansion
coefficient for the NHM function $\mathbf{U}_{n}(\mathbf{R)}$, whilst $P_{n}$
is that for the HAM function $\mathbf{V}_{n}(\mathbf{R)}$. This role
reversal for the generalised momenta is a consequence of introducing the
bi-orthogonal NHM functions.

\subsection{\textit{Hamiltonian}}

The Hamiltonian $H$ is defined via the true mode expression and obtained in
terms of true mode generalised coordinates and momenta: 
\begin{eqnarray}
H &=&\sum_{k}^{R}\,[p_{k}\dot{q}_{k}^{\ast }+p_{k}^{\ast }\dot{q}_{k}]-L
\label{H.def} \\
H &=&\sum_{k}\,(p_{k}p_{k}^{\ast }+\epsilon _{0}^{2}\omega
_{k}^{2}q_{k}q_{k}^{\ast })/2\epsilon _{0}.  \label{Htrue.eq}
\end{eqnarray}
The Hamiltonian is that for independent harmonic oscillators, one for each
true mode in each half space. To obtain the Hamiltonian in terms of NHM we
first write the Hamitonian as the sum of cavity and external contributions $%
H_{C},H_{E}$ 
\begin{equation}
H=H_{C}+H_{E}  \label{Hsum.eq}
\end{equation}
given by 
\begin{eqnarray}
H_{C} &=&\sum_{k}^{R}\,[p_{k}^{C}\dot{q}_{k}^{\ast }+p_{k}^{C\ast }\dot{q}%
_{k}]-L_{C}  \label{Hc.def} \\
H_{E} &=&\sum_{k}^{R}\,[p_{k}^{E}\dot{q}_{k}^{\ast }+p_{k}^{E\ast }\dot{q}%
_{k}]-L_{E}.  \label{He.def}
\end{eqnarray}
We then show that 
\begin{eqnarray}
p_{k}^{C} &=&\sum_{n}(\mathbf{\Lambda }_{nk}^{{\ast }}\,\dot{Q}_{n}+\mathbf{%
\Gamma }_{nk}^{{\ast }}\,\dot{R}_{n})/2\epsilon _{0}  \label{pCk.eq} \\
p_{k}^{E} &=&\sum_{K}(\mathbf{\Phi }_{Kk}^{{\ast }}\,\dot{Q}_{K}+\mathbf{%
\Delta }_{Kk}^{{\ast }}\,\dot{R}_{K})/2\epsilon _{0}.  \label{pEk.eq}
\end{eqnarray}
Equations (\ref{QRq.eq}) etc. then enable us to eliminate the $k$ sums
involving $\dot{q}_{k},\dot{q}_{k}^{\ast }$, so that $H_{C},H_{E}$ now only
contain NHM generalised velocities $\dot{Q}_{n},\dot{R}_{n},\dot{Q}_{K},\dot{%
R}_{n}$. Eliminating all the NHM\ generalised velocities in favour of the
NHM\ generalised momenta via $P_{n}=\epsilon _{0}\dot{R}_{n}$, $%
S_{n}=\epsilon _{0}\dot{Q}_{n}$ etc. finally gives the contributions to the
Hamiltonian $H_{C},H_{E}$ from the cavity and external regions. The cavity
contribution $H_{C}$ only depends on $Q_{n},P_{n}$ or $R_{n},S_{n}$. and the
external contribution $H_{E}$ only depends on $Q_{K},P_{K}$ or $R_{K},S_{K}$%
. Expressed in terms of just one set of generalised coordinates and momenta
(the $Q,P$ or the $R,S$ choice), both $H_{C}$ and $H_{E}$ are Hamiltonians
for coupled harmonic oscillators, \textit{two} oscillators for each $n$ and $%
K$ and there will be double sums over $n$ and $K$. However if a mixture of
both choices are used, the cavity and external region contributions to the
Hamiltonian can be expressed as single sums:

\begin{eqnarray}
H_{C} &=&\sum_{n}\,([P_{n}{}^{{\large \ast }}S_{n}+S_{n}{}^{{\large \ast }%
}P_{n}]+\epsilon _{0}^{2}\omega _{n}^{2}[Q_{n}\,R_{n}{}^{{\large \ast }%
}+R_{n}\,Q_{n}{}^{{\large \ast }}])/2\epsilon _{0}  \label{Hc.eq} \\
H_{E} &=&\sum_{K}\,([P_{K}{}^{{\large \ast }}S_{K}+S_{K}{}^{{\large \ast }%
}P_{K}]+\epsilon _{0}^{2}\omega _{K}^{2}[Q_{K}\,R_{K}{}^{{\large \ast }%
}+R_{K}\,Q_{K}{}^{{\large \ast }}])/2\epsilon _{0}.  \notag \\
&&  \label{He.eq}
\end{eqnarray}
Using the transformation relations in equations (\ref{RQ.eq}, \ref{PiT.eq}),
the sums over $n$ of the two terms in each of the square brackets in
equation (\ref{Hc.eq}) can be shown to be equal, however we have chosen to
leave the Hamiltonian in a more symmetrical form regarding the two sets of
generalised position and momentum coordinates.

\subsection{\textit{Canonical quantization and commutation rules}}

Canonical quantization involves replacing the generalised coordinates and
momenta by quantum operators satisfying standard commutation rules. The
replacements are: 
\begin{equation}
q_{k}\rightarrow \hat{q}_{k},\quad q_{k}^{\ast }\rightarrow \hat{q}%
_{k}{}^{\dagger }=\hat{q}_{k\ast }\quad ,p_{k}\rightarrow \hat{p}_{k},\quad
p_{k}^{\ast }\rightarrow \hat{p}_{k}{}^{\dagger }=\hat{p}_{k\ast }
\label{Canon1.eq}
\end{equation}
for true modes and 
\begin{eqnarray}
Q_{n} &\rightarrow &\hat{Q}_{n},\quad (Q_{n})^{\ast }\rightarrow \hat{Q}%
_{n}{}^{\dagger }\quad ,P_{n}\rightarrow \hat{P}_{n},\quad (P_{n})^{\ast
}\rightarrow \hat{P}_{n}{}^{\dagger },  \notag \\
R_{n} &\rightarrow &\hat{R}_{n},\quad (R_{n})^{\ast }\rightarrow \hat{R}%
_{n}{}^{\dagger }\quad ,S_{n}\rightarrow \hat{S}_{n},\quad (S_{n})^{\ast
}\rightarrow \hat{S}_{n}{}^{\dagger }  \label{Canon2.eq}
\end{eqnarray}
for the cavity NHM. The standard commutators that are non zero are: \ 
\begin{equation}
\lbrack \hat{q}_{k},\hat{p}_{l}{}^{\dagger }]=[\hat{q}_{k}{}^{\dagger },\hat{%
p}_{l}]=i\hbar \,\delta _{kl}  \label{pkqkCR.eq}
\end{equation}
from which the commutators for the cavity NHM operators can be obtained via
the relationshps in equations (\ref{QRq.eq}, \ref{PiTp.eq}) and the
properties for the $\mathbf{\Gamma },\mathbf{\Lambda }$ matrices. For the
cavity NHM operators the non zero results are: 
\begin{equation}
\lbrack \hat{Q}_{n},\hat{P}_{m}{}^{\dagger }]=[\hat{Q}_{n}{}^{\dagger },\hat{%
P}_{m}]=[\hat{R}_{n},\hat{S}_{m}{}^{\dagger }]=[\hat{R}_{n}{}^{\dagger },%
\hat{S}_{m}]=i\hbar \,\,\delta _{nm}  \label{QnPnRnSnCR.eq}
\end{equation}
and 
\begin{eqnarray}
\lbrack \hat{Q}_{n},\hat{S}_{m}{}^{\dagger }] &=&i\hbar \,\,\mathbf{D}%
_{nm}\qquad \;[\hat{Q}_{n}{}^{\dagger },\hat{S}_{m}]=i\hbar \,\mathbf{D}_{mn}
\label{QnSnCR.eq} \\
\lbrack \hat{R}_{n},\hat{P}_{m}{}^{\dagger }] &=&i\hbar \,\,\mathbf{C}%
_{nm}\qquad \;[\hat{R}_{n}{}^{\dagger },\hat{P}_{m}]=i\hbar \,\,\mathbf{C}%
_{mn}.  \label{RnPnCr.eq}
\end{eqnarray}
$\;$ Similar replacements are made for the external NHM generalised
coordinates and momenta, and commutation rules analogous to those for the $%
\hat{Q}_{n},\hat{P}_{n},\hat{R}_{n},\hat{S}_{n}$ are obtained for the $\hat{Q%
}_{K},\hat{P}_{K},\hat{R}_{K},\hat{S}_{K}$. The commutators involving both
cavity and external NHM operators will be discussed below.\textit{\ }The
Hamiltonian $H$ becomes the operator $\hat{H}$ and the vector potential and
conjugate momentum fields $\mathbf{A(R),\Pi (R)}$ become field operators $%
\mathbf{\hat{A}(R),\hat{\Pi}(R)}$. The formal expressions for these
operators in terms of the generalised coordinate and momentum operators are
exactly the same as in the classical expressions, so need not be repeated
here.

The commutation rules for the field operators $\mathbf{\hat{A}(R),\hat{\Pi}%
(R)}$ can be obtained via their true mode expressions and are given by:

\begin{equation}
\lbrack \mathbf{\hat{A}}^{\alpha }\mathbf{(R),\hat{\Pi}}^{\beta }\mathbf{(R}%
^{/}\mathbf{)]=\,}i\hbar \,\delta _{\alpha \beta }\,\delta (\mathbf{R-R}%
^{/}).  \label{APiCR.eq}
\end{equation}
Those for the electric displacement field operator $\mathbf{\hat{D}(R)=-\hat{%
\Pi}(R)}$ and the magnetic field operator $\mathbf{\hat{B}(R)=\nabla \times 
\hat{A}(R)}$ can then be obtained as:

\begin{equation}
\lbrack \mathbf{\hat{B}}^{\alpha }\mathbf{(R),\hat{D}}^{\beta }\mathbf{(R}%
^{/}\mathbf{)]=\,}i\hbar \,\sum_{\gamma }\varepsilon _{\alpha \beta \gamma
}\,\partial /\partial X_{\gamma }\,\delta (\mathbf{R-R}^{/})  \label{BDCR.eq}
\end{equation}

The commutators involving both cavity and external NHM operators are less
straightforward to determine than those just involving one type at a time.
One suitable approach involves using the commutation rules obtained for the
electric displacement and magnetic field operators. The details of this
calculation are of some interest, and are set out in the Appendix. The non
zero results can be expressed in terms of surface integrals over the
boundary $S$\ between the cavity and external regions, and are given below
as $I_{nK},J_{nK},K_{nK},L_{nK},\mathcal{I}_{nK},\mathcal{J}_{nK},\mathcal{K}%
_{nK},\mathcal{L}_{nK}$. Apart from others obtained by taking the Hermitean
adjoint, the non zero commutators are\textit{\ }for the case of cavity
generalised coordinates and external generalised momenta are: 
\begin{eqnarray}
\lbrack \hat{Q}_{n},\hat{P}_{K}^{\dagger }] &=&(\hbar
\,/\,k_{n})\;\int_{S}\,d^{2}\mathbf{s\,V}_{n}^{\ast }\cdot \mathbf{U}%
_{K}=i\hbar \,L_{nK}  \label{CEcr1.eq} \\
\lbrack \hat{R}_{n},\hat{S}_{K}^{\dagger }] &=&(\hbar
\,/\,k_{n})\;\int_{S}\,d^{2}\mathbf{s\,U}_{n}^{\ast }\cdot \mathbf{V}%
_{K}=i\hbar \,K_{nK}  \label{CEcr2.eq} \\
\lbrack \hat{Q}_{n},\hat{S}_{K}^{\dagger }] &=&(\hbar
\,/\,k_{n})\;\int_{S}\,d^{2}\mathbf{s\,V}_{n}^{\ast }\cdot \mathbf{V}%
_{K}=i\hbar \,J_{nK}  \label{CEcr3.eq} \\
\lbrack \hat{R}_{n},\hat{P}_{K}^{\dagger }] &=&(\hbar
\,/\,k_{n})\;\int_{S}\,d^{2}\mathbf{s\,U}_{n}^{\ast }\cdot \mathbf{U}%
_{K}=i\hbar \,I_{nK}  \label{CEcr4.eq}
\end{eqnarray}
The non zero commutators for the case of cavity generalised momenta and
external generalised coordinates are:

\begin{eqnarray}
\lbrack \hat{P}_{n},\hat{Q}_{K}^{\dagger }] &=&(i\,\hbar
\,/\,k_{K}^{2})\;\int_{S}\,d^{2}\mathbf{s\;\hat{z}\cdot U}_{n}^{\ast }\times
(\mathbf{\nabla \times V}_{K})=-i\hbar \,\mathcal{K}_{nK}  \label{CEcr5.eq}
\\
\lbrack \hat{S}_{n},\hat{R}_{K}^{\dagger }] &=&(i\,\hbar
\,/\,k_{K}^{2})\;\int_{S}\,d^{2}\mathbf{s\;\hat{z}\cdot V}_{n}^{\ast }\times
(\mathbf{\nabla \times U}_{K})=-i\hbar \,\mathcal{L}_{nK}  \label{CEcr6.eq}
\\
\lbrack \hat{P}_{n},\hat{R}_{K}^{\dagger }] &=&(i\,\hbar
\,/\,k_{K}^{2})\;\int_{S}\,d^{2}\mathbf{s\;\hat{z}\cdot U}_{n}^{\ast }\times
(\mathbf{\nabla \times U}_{K})=-i\hbar \,\mathcal{I}_{nK}  \label{CEcr7.eq}
\\
\lbrack \hat{S}_{n},\hat{Q}_{K}^{\dagger }] &=&(i\,\hbar
\,/\,k_{K}^{2})\;\int_{S}\,d^{2}\mathbf{s\;\hat{z}\cdot V}_{n}^{\ast }\times
(\mathbf{\nabla \times V}_{K})=-i\hbar \,\mathcal{J}_{nK}  \label{CEcr8.eq}
\end{eqnarray}
These commutators are much smaller than those involving only cavity
quantities or only external quantities.

\subsection{\textit{Annihilation and creation operators}}

In view of the link between the various classical Hamiltonian terms and
harmonic oscillators, it is appropriate to replace the generalised
coordinate and momentum operators by annihilation and creation operators .
For the true modes we define:

\begin{eqnarray}
\hat{a}_{k} &=&\lambda _{k}\hat{q}_{k}+i\mu _{k}\hat{p}_{k}\qquad \hat{a}%
_{k}^{{\large \dagger }}=\lambda _{k}\hat{q}_{k}^{{\large \dagger }}-i\mu
_{k}\hat{p}_{k}^{{\large \dagger }}  \notag \\
\hat{a}_{k\ast } &=&\lambda _{k}\hat{q}_{k}^{{\large \dagger }}+i\mu _{k}%
\hat{p}_{k}^{{\large \dagger }}\qquad \hat{a}_{k\ast }^{{\large \dagger }%
}=\lambda _{k}\hat{q}_{k}-i\mu _{k}\hat{p}_{k},  \label{annihtrue.eq}
\end{eqnarray}
where $\lambda _{k}=\sqrt{\epsilon _{0}\omega _{k}/2\hbar }$ and $\mu _{k}=%
\sqrt{1/2\hbar \epsilon _{0}\omega _{k}}$. These have non zero commutation
rules $[\hat{a}_{k},\hat{a}_{l}^{\dagger }]=\delta _{kl}$ and the
Hamiltonian operator can be written as the well known form 
\begin{equation}
\hat{H}=\sum_{k}\,(\hat{a}_{k}^{\dagger }\hat{a}_{k}+1/2).\hbar \omega _{k},
\label{HtrueQuant.eq}
\end{equation}
corresponding to independent quantum harmonic oscillators, one for each true
mode in both half spaces. For the NHM a somewhat unusual choice is made
owing to the juxtaposition of the $\hat{Q},\hat{R}$ operators and $\hat{P},%
\hat{S}$ operators in the Hamiltonian contributions $\hat{H}_{C},\hat{H}_{E}$%
. We combine the $\hat{Q}$ with the $\hat{S}$ and the $\hat{R}$ with the $%
\hat{P}$ along with the Hermitean adjoints to define annihilation and
creation operators.

For the cavity NHM it is possible to define eight operators of the
annihilation and creation operator type:

\begin{eqnarray}
\hat{A}_{n} &=&\lambda _{n}\hat{Q}_{n}+i\mu _{n}\hat{S}_{n}\qquad \qquad 
\hat{A}_{n}^{{\Large \#}}=\lambda _{n}\hat{R}_{n}^{{\Large \dagger }}-i\mu
_{n}\hat{P}_{n}^{{\Large \dagger }}  \notag \\
\hat{B}_{n} &=&\lambda _{n}\hat{R}_{n}^{{\Large \dagger }}+i\mu _{n}\hat{P}%
_{n}^{{\Large \dagger }}\qquad \qquad \hat{B}_{n}^{{\Large \#}}=\lambda _{n}%
\hat{Q}_{n}-i\mu _{n}\hat{S}_{n}  \notag \\
\hat{A}_{n}^{{\Large \dagger }} &=&\lambda _{n}\hat{Q}_{n}^{{\Large \dagger }%
}-i\mu _{n}\hat{S}_{n}^{{\Large \dagger }}\qquad \qquad \hat{A}_{n}^{\#%
{\Large \dagger }}=\lambda _{n}\hat{R}_{n}{}+i\mu _{n}\hat{P}_{n}  \notag \\
\hat{B}_{n}^{{\Large \dagger }} &=&\lambda _{n}\hat{R}_{n}{}-i\mu _{n}\hat{P}%
_{n}{}\qquad \qquad \hat{B}_{n}^{\#{\Large \dagger }}=\lambda _{n}\hat{Q}%
_{n}^{{\Large \dagger }}+i\mu _{n}\hat{S}_{n}^{{\Large \dagger }},
\label{annihNHM.eq}
\end{eqnarray}
which are expressed as linear combinations of the eight original generalised
position and momentum operators and their adjoints. With these choices we
see that the annihilation, creation operator pairs are now $(\hat{A}_{n},%
\hat{A}_{n}^{\#})$, $(\hat{B}_{n},\hat{B}_{n}^{\#})$, $(\hat{A}%
_{n}^{\#\dagger },\hat{A}_{n}^{\dagger })$ and $(\hat{B}_{n}^{\#\dagger },%
\hat{B}_{n}^{\dagger })$ rather than the expected $(\hat{A}_{n},\hat{A}%
_{n}^{\dagger })$, $(\hat{B}_{n},\hat{B}_{n}^{\dagger })$, $(\hat{A}%
_{n}^{\#\dagger },\hat{A}_{n}^{\#})$ and $(\hat{B}_{n}^{\#\dagger },\hat{B}%
_{n}^{\#})$. The non zero commutation rules are: 
\begin{equation}
\lbrack \hat{A}_{n},\hat{A}_{m}^{\#}]=[\hat{B}_{n},\hat{B}_{m}^{\#}]=[\hat{A}%
_{n}^{\#\dagger },\hat{A}_{m}^{\dagger }]=[\hat{B}_{n}^{\#\dagger },\hat{B}%
_{m}^{\dagger }]=\delta _{nm}  \label{CavNHMAnnCreatCR.eq}
\end{equation}
and 
\begin{eqnarray}
\lbrack \hat{A}_{n},\hat{A}_{m}^{\dagger }] &=&\,\mathbf{D}_{nm}\qquad
\lbrack \hat{B}_{n}^{\#\dagger },\hat{B}_{m}^{\#}]=\,\mathbf{D}_{mn}
\label{CavNHMAnnCreatCR2.eq} \\
\lbrack \hat{B}_{n},\hat{B}_{m}^{\dagger }] &=&\,\mathbf{C}_{mn}\qquad
\lbrack \hat{A}_{n}^{\#\dagger },\hat{A}_{m}^{\#}]=\,\mathbf{C}_{nm}.
\label{CavNHMAnnCreatCR3.eq}
\end{eqnarray}
Some commutators are zero because of the selection rule $\omega _{n}=\omega
_{m}$ for non zero $\mathbf{C}_{nm}$ and $\mathbf{D}_{nm}$. Similar
commutation rules occur in reference \cite{Lamprecht99} (their $\hat{a}_{n}$
corresponds to our $\hat{A}_{n}$, their $\hat{b}_{n}^{\dagger }$ to our $%
\hat{A}_{n}^{\#}$), except that here we have twice as many annihilation,
creation operator pairs since no operators corresponding to our $\hat{B}_{n}$
and $\hat{B}_{n}^{\#}$ occur. The precise nature of the eight operators can
be seen by expressing each of them as a linear combination of the true mode
annihilation and creation operators. Each turns out to actually involve both
types of true mode operator, for example $\hat{A}_{n}$ is a linear
combination of the $\hat{a}_{k}$ and the $\hat{a}_{k\ast }^{\dagger }$ with $%
k$ in the right half space. However, the expansion coefficient for one type
of operator involves the factor $f_{-}(\omega _{n},\omega _{k})=(\omega
_{n}-\omega _{k})/(2\sqrt{\omega _{n}\omega _{k}})$, the other the factor $%
f_{+}(\omega _{n},\omega _{k})=(\omega _{n}+\omega _{k})/(2\sqrt{\omega
_{n}\omega _{k}})$. Assuming that all the field modes of interest have
approximately the same frequency (monochromaticity approximation, cf. \cite
{Lamprecht99}) terms involve the first factor can be ignored and the second
factor approximated as unity. We then find the approximate relationships:

\begin{eqnarray}
\hat{A}_{n} &\cong &\sum_{k}^{R}\mathbf{\Gamma }_{nk}\,\hat{a}_{k}\qquad 
\hat{A}_{n}^{{\Large \#}}\cong \sum_{k}^{R}\mathbf{\Lambda }_{nk}^{{\large %
\ast }}\,\hat{a}_{k}^{{\large \dagger }}  \notag \\
\hat{B}_{n} &\cong &\sum_{k}^{R}\mathbf{\Lambda }_{nk}^{{\large \ast }}\,%
\hat{a}_{k\ast }\qquad \hat{B}_{n}^{{\Large \#}}\cong \sum_{k}^{R}\mathbf{%
\Gamma }_{nk}\,\hat{a}_{k\ast }^{{\large \dagger }}  \notag \\
\hat{A}_{n}^{\#{\Large \dagger }} &\cong &\sum_{k}^{R}\mathbf{\Lambda }%
_{nk}\,\hat{a}_{k}\qquad \hat{A}_{n}^{{\Large \dagger }}\cong \sum_{k}^{R}%
\mathbf{\Gamma }_{nk}^{{\large \ast }}\,\hat{a}_{k}^{{\large \dagger }} 
\notag \\
\hat{B}_{n}^{\#{\Large \dagger }} &\cong &\sum_{k}^{R}\mathbf{\Gamma }_{nk}^{%
{\large \ast }}\,\hat{a}_{k\ast }\qquad \hat{B}_{n}^{{\Large \dagger }}\cong
\sum_{k}^{R}\mathbf{\Lambda }_{nk}\,\,\hat{a}_{k\ast }^{{\large \dagger }}.
\label{approxNHMtrue.eq}
\end{eqnarray}
We thus can identify $\hat{A}_{n}$ and $\hat{B}_{n}$ as annihilation
operators associated with right and left travelling modes respectively, and
with $\hat{A}_{n}^{\#}$ and $\hat{B}_{n}^{\#}$ as their corresponding
creation operators. A similar identification of $\hat{A}_{n}^{\#\dagger }$
and $\hat{B}_{n}^{\#\dagger }$ as annihilation operators, $\hat{A}%
_{n}^{\dagger }$ and $\hat{B}_{n}^{\dagger }$ the corresponding creation
operators also applies. Using the selection rule $\omega _{n}=\omega _{m}$
for non zero $\mathbf{C}_{nm}$ and $\mathbf{D}_{nm}$, the eight operators
for the cavity NHM can be interrelated as follows:

\begin{eqnarray}
\hat{A}_{n}^{\#{\Large \dagger }} &=&\sum_{m}\mathbf{C}_{nm}\,\hat{A}%
_{m}\qquad \hat{B}_{n}^{{\Large \dagger }}=\sum_{m}\mathbf{C}_{nm}\,\hat{B}%
_{m}^{{\Large \#}}  \notag \\
\hat{A}_{n}^{{\Large \dagger }} &=&\sum_{m}\mathbf{D}_{mn}\,\hat{A}_{m}^{%
{\Large \#}}\qquad \hat{B}_{n}^{\#{\Large \dagger }}=\sum_{m}\mathbf{D}%
_{mn}\,\hat{B}_{m}.  \label{CavNHMrel.eq}
\end{eqnarray}
These relationships are used to simplify the cavity term in the Hamiltonian $%
\hat{H}_{C}$.

A similar approach can be used to construct annihilation and creation
operators for the external NHM. For the external NHM we write:

\begin{eqnarray}
\hat{A}_{K} &=&\lambda _{K}\hat{Q}_{K}+i\mu _{K}\hat{S}_{K}\qquad \qquad 
\hat{A}_{K}^{{\Large \#}}=\lambda _{K}\hat{R}_{K}^{{\Large \dagger }}-i\mu
_{K}\hat{P}_{K}^{{\Large \dagger }}  \notag \\
\hat{B}_{K} &=&\lambda _{K}\hat{R}_{K}^{{\Large \dagger }}+i\mu _{K}\hat{P}%
_{K}^{{\Large \dagger }}\qquad \qquad \hat{B}_{K}^{{\Large \#}}=\lambda _{K}%
\hat{Q}_{K}-i\mu _{K}\hat{S}_{K}  \notag \\
\hat{A}_{K}^{{\Large \dagger }} &=&\lambda _{K}\hat{Q}_{K}^{{\Large \dagger }%
}-i\mu _{K}\hat{S}_{K}^{{\Large \dagger }}\qquad \qquad \hat{A}_{K}^{\#%
{\Large \dagger }}=\lambda _{K}\hat{R}_{K}{}+i\mu _{K}\hat{P}_{K}  \notag \\
\hat{B}_{K}^{{\Large \dagger }} &=&\lambda _{K}\hat{R}_{K}{}-i\mu _{K}\hat{P}%
_{K}{}\qquad \qquad \hat{B}_{K}^{\#{\Large \dagger }}=\lambda _{K}\hat{Q}%
_{K}^{{\Large \dagger }}+i\mu _{K}\hat{S}_{K}^{{\Large \dagger }}.
\label{annihextNHM.eq}
\end{eqnarray}
The commutation rules are analogous to those for the cavity NHM operators,
except that some further commutators are non zero because the selection rule 
$\omega _{K}=\omega _{L}$ for non zero $\mathbf{G}_{KL}$ and $\mathbf{H}%
_{KL} $ is only approximate, and the expected results like $[\hat{A}_{K},%
\hat{A}_{L}^{\dagger }]=\,\mathbf{H}_{KL}$ are multiplied by the factor $%
f_{+}(\omega _{K},\omega _{L})$ that is only approximately unity.
Approximate relations between the external NHM annihilation and creation
operators and the true mode operators that are analogous to equations (\ref
{approxNHMtrue.eq}) and interrelations between the external NHM operators
that are analogous to equations (\ref{CavNHMrel.eq}) can be obtained. These
depend respectively, on the monochromaticity assumption and the approximate
selection rules on $\mathbf{G}_{KL},\mathbf{H}_{KL}$. Operators such as $%
\hat{A}_{K}$ and $\hat{B}_{K}$ are approximate linear combinations of true
mode annihilation operators, with $\hat{A}_{K}^{\#}$ and $\hat{B}_{K}^{\#}$
their corresponding creation operators.

\subsection{\textit{Field operators}}

The field operators $\mathbf{\hat{A}(R),\hat{\Pi}(R)}$ can be now expressed
in terms of annihilation and creation operators. In terms of true modes we
have:

\begin{eqnarray}
\mathbf{\hat{A}(R)} &=&\sum_{k}^{R}\sqrt{\hbar \,/\,2\epsilon _{0}\omega _{k}%
}([\hat{a}_{k}+\hat{a}_{k\ast }^{{\Large \dagger }}]\,\mathbf{A}_{k}(\mathbf{%
R)+}[\hat{a}_{k}^{{\Large \dagger }}+\hat{a}_{k\ast }]\,\mathbf{A}_{k}(%
\mathbf{R)}^{{\Large \ast }}\mathbf{)}  \label{Atrue1.eq} \\
&=&\sum_{k}\sqrt{\hbar \,/\,2\epsilon _{0}\omega _{k}}(\hat{a}_{k}\,\mathbf{A%
}_{k}(\mathbf{R)+}\hat{a}_{k}^{{\Large \dagger }}\,\mathbf{A}_{k}(\mathbf{R)}%
^{{\Large \ast }}\mathbf{)}  \label{Atrue2.eq}
\end{eqnarray}
and 
\begin{eqnarray}
\mathbf{\hat{\Pi}(R)} &=&\sum_{k}^{R}\frac{1}{i}\sqrt{\hbar \epsilon
_{0}\omega _{k}\,/\,2}([\hat{a}_{k}-\hat{a}_{k\ast }^{{\Large \dagger }}]\,%
\mathbf{A}_{k}(\mathbf{R)-}[\hat{a}_{k}^{{\Large \dagger }}-\hat{a}_{k\ast
}]\,\mathbf{A}_{k}(\mathbf{R)}^{{\Large \ast }}\mathbf{)}  \label{PiTRUE1.eq}
\\
&=&\sum_{k}\frac{1}{i}\sqrt{\hbar \epsilon _{0}\omega _{k}\,/\,2}(\hat{a}%
_{k}\,\mathbf{A}_{k}(\mathbf{R)-}\hat{a}_{k}^{{\Large \dagger }}\,\mathbf{A}%
_{k}(\mathbf{R)}^{{\Large \ast }}\mathbf{)}  \label{PITRUE2.eq}
\end{eqnarray}
for all positions $\mathbf{R}$. Expanding in NHM functions the field
operators are:

\begin{eqnarray}
\mathbf{\hat{A}(R)} &=&\sum_{n}\sqrt{\hbar \,/\,2\epsilon _{0}\omega _{n}}([%
\hat{A}_{n}+\hat{B}_{n}^{{\Large \#}}]\,\mathbf{U}_{n}(\mathbf{R)+}[\hat{A}%
_{n}^{{\Large \#}}+\hat{B}_{n}]\,\mathbf{V}_{n}(\mathbf{R)}^{{\Large \ast }}%
\mathbf{)}  \label{ANHM.eq} \\
\mathbf{\hat{\Pi}(R)} &=&\sum_{n}\frac{1}{i}\sqrt{\hbar \epsilon _{0}\omega
_{n}\,/\,2}([\hat{A}_{n}-\hat{B}_{n}^{{\Large \#}}]\,\mathbf{U}_{n}(\mathbf{%
R)-}[\hat{A}_{n}^{{\Large \#}}-\hat{B}_{n}]\,\mathbf{V}_{n}(\mathbf{R)}^{%
{\Large \ast }}\mathbf{),}  \label{PINHM.eq}
\end{eqnarray}
for positions $\mathbf{R}$ in the cavity region. The monochromaticity
approximation is not required. Comparing equations (\ref{Atrue1.eq}, \ref
{PiTRUE1.eq}) with equations (\ref{ANHM.eq}, \ref{PINHM.eq}) again suggests
the identification of $\hat{A}_{n}$ and $\hat{B}_{n}$ as annihilation
operators associated with right and left travelling modes respectively, and
also indicates that we have twice as many annihilation operators as in
reference \cite{Lamprecht99}. Their expressions corresponding to equations (%
\ref{ANHM.eq}, \ref{PINHM.eq}) do not include the $\hat{B}_{n}^{\#}$ and $%
\hat{B}_{n}$ contributions.

\subsection{\textit{Quantum Hamiltonian}}

The cavity term in the Hamiltonian can be expressed in terms of the cavity
NHM annihilation and creation operators. We find that:

\begin{eqnarray}
\hat{H}_{C} &=&\sum_{n}\,\hbar \omega _{n}\{\,(\frac{1}{2}[\hat{A}_{n}^{%
{\Large \#}}\hat{A}_{n}+\hat{A}_{n}^{{\Large \dagger }}\hat{A}_{n}^{\#%
{\Large \dagger }}]+\frac{1}{2})+\,(\frac{1}{2}[\hat{B}_{n}^{{\Large \#}}%
\hat{B}_{n}+\hat{B}_{n}^{{\Large \dagger }}\hat{B}_{n}^{\#{\Large \dagger }}%
]+\frac{1}{2})\}  \notag \\
&=&\sum_{n}\,\hbar \omega _{n}\{\,(\hat{A}_{n}^{{\Large \dagger }}\hat{A}%
_{n}^{\#{\Large \dagger }}+\frac{1}{2})+\,(\hat{B}_{n}^{{\Large \dagger }}%
\hat{B}_{n}^{\#{\Large \dagger }}+\frac{1}{2})\}  \label{Hc2Dag.eq} \\
&=&\sum_{n}\,\hbar \omega _{n}\{\,(\hat{A}_{n}^{{\Large \#}}\hat{A}_{n}+%
\frac{1}{2})+\,(\hat{B}_{n}^{{\Large \#}}\hat{B}_{n}+\frac{1}{2})\}.
\label{Hc2.eq}
\end{eqnarray}
The first form is obtained just by inverting equations (\ref{annihNHM.eq})
and substituting into the operator form of equation (\ref{Hc.eq}). The
derivation of the second and third forms for $\hat{H}_{C}$ uses equations (%
\ref{CavNHMrel.eq}), which is based on the selection rule $\omega
_{n}=\omega _{m}$ for non zero $\mathbf{C}_{nm}$ and $\mathbf{D}_{nm}$. The
final expression for the cavity contribution $\hat{H}_{C}$ to the
Hamiltonian turns out to be the Hamiltonian for a set of independent quantum
harmonic oscillators (QHO), two for each cavity NHM index $n$ - one
corresponding to the left travelling NHM, the other to the right travelling
NHM.. The lack of coupling between the cavity NHM QHO follows from the
result that all $\hat{A}_{n}^{\#}\hat{A}_{n}$ and $\hat{B}_{n}^{\#}\hat{B}%
_{n}$ commute with each other. Because of the commutation rules, the
quantities $\hat{A}_{n}^{\#}\hat{A}_{n}$\ and $\hat{B}_{n}^{\#}\hat{B}_{n}$\
act as number operators, even though they are not Hermitean. However,
although the individual terms in equation (\ref{Hc2.eq}) are not Hermitean,
the overall sum giving $\hat{H}_{C}$\ is. The existence of QHO will become
apparent when the approximate energy eigenstates for the Hamiltonian $\hat{H}
$ are obtained.

Expressions for the field operators in the external region analogous to
equations (\ref{ANHM.eq}, \ref{PINHM.eq}) apply however without using the
approximate relations between the external NHM annihilation and creation
operators and the true mode operators that are analogous to equations (\ref
{approxNHMtrue.eq}) and the interrelationships between the external NHM
operators that are analogous to equations (\ref{CavNHMrel.eq}), and more
importantly the external region contribution to the Hamiltonian $\hat{H}_{E}$
can be written as:

\begin{equation}
\hat{H}_{E}=\sum_{K}\,\hbar \omega _{K}\{\,(\frac{1}{2}[\hat{A}_{K}^{{\Large %
\#}}\hat{A}_{K}+\hat{A}_{K}^{{\Large \dagger }}\hat{A}_{K}^{\#{\Large %
\dagger }}]+\frac{1}{2})+\,(\frac{1}{2}[\hat{B}_{K}^{{\Large \#}}\hat{B}_{K}+%
\hat{B}_{K}^{{\Large \dagger }}\hat{B}_{K}^{\#{\Large \dagger }}]+\frac{1}{2}%
)\}  \label{He2.eq}
\end{equation}
If we now apply the relationships that analogous to equations (\ref
{CavNHMrel.eq}), the last expression can be simplified to give an
approximate form for $\hat{H}_{E}$:

\begin{equation}
\hat{H}_{E}^{0}=\sum_{K}\,\hbar \omega _{K}\{\,(\hat{A}_{K}^{{\Large \#}}%
\hat{A}_{K}+\frac{1}{2})+\,(\hat{B}_{K}^{{\Large \#}}\hat{B}_{K}+\frac{1}{2}%
)\}.  \label{He0.eq}
\end{equation}
The approximate external region contribution $\hat{H}_{E}^{0}$ to the field
Hamiltonian will be seen to be the Hamiltonian for a set of independent
QHOs, two for each external region NHM $K$. Independence follows from all
the $\hat{A}_{K}^{\#}\hat{A}_{K}$ and $\hat{B}_{K}^{\#}\hat{B}_{K}$ factors
commuting with each other. Again, the non Hermitean quantities $\hat{A}%
_{K}^{\#}\hat{A}_{K}$\ and $\hat{B}_{K}^{\#}\hat{B}_{K}$\ act as number
operators\textit{.} The full expression for the external region contribution
to the Hamiltonian can then be written as the sum of the harmonic oscillator
term $\hat{H}_{E}^{0}$ and a coupling term $\hat{V}_{E}$ as follows: 
\begin{equation}
\hat{H}_{E}=\hat{H}_{E}^{0}+\hat{V}_{E}  \label{QuantHeSum.eq}
\end{equation}
where the coupling term $\hat{V}_{E}$ is

\begin{equation}
\hat{V}_{E}=\frac{1}{2}\sum_{K}\,\hbar \omega _{K}\{\,\hat{A}_{K}^{{\Large %
\dagger }}\hat{A}_{K}^{\#{\Large \dagger }}+\hat{B}_{K}^{{\Large \dagger }}%
\hat{B}_{K}^{\#{\Large \dagger }}\}-\,\frac{1}{2}\sum_{K}\hbar \omega
_{K}\{\,\hat{A}_{K}^{{\Large \#}}\hat{A}_{K}+\,\hat{B}_{K}^{{\Large \#}}\hat{%
B}_{K}\}  \label{Ve.eq}
\end{equation}
It should be noted that neither $\hat{H}_{E}^{0}$ nor $\hat{V}_{E}$ are
Hermitean, though their sum giving $\hat{H}_{E}$ is. It is also possible to
write: 
\begin{equation}
\hat{H}_{E}=(\hat{H}_{E}^{0})^{\dagger }+(\hat{V}_{E})^{\dagger },
\label{QuantHeSum2.eq}
\end{equation}
where: 
\begin{eqnarray}
(\hat{H}_{E}^{0})^{\dagger } &=&\sum_{K}\,\hbar \omega _{K}\{\,(\hat{A}_{K}^{%
{\Large \dagger }}\hat{A}_{K}^{\#{\Large \dagger }}+\frac{1}{2})+\,(\hat{B}%
_{K}^{{\Large \dagger }}\hat{B}_{K}^{\#{\Large \dagger }}+\frac{1}{2})\}
\label{He0Dag.eq} \\
(\hat{V}_{E})^{\dagger } &=&\frac{1}{2}\sum_{K}\,\hbar \omega _{K}\{\,\hat{A}%
_{K}^{{\Large \#}}\hat{A}_{K}+\hat{B}_{K}^{{\Large \#}}\hat{B}_{K}\}  \notag
\\
&&-\,\frac{1}{2}\sum_{K}\hbar \omega _{K}\{\,\hat{A}_{K}^{{\Large \dagger }}%
\hat{A}_{K}^{\#{\Large \dagger }}+\,\hat{B}_{K}^{{\Large \dagger }}\hat{B}%
_{K}^{\#{\Large \dagger }}\}  \label{VeDag.eq}
\end{eqnarray}

The field Hamiltonian $\hat{H}$\ is now given by: 
\begin{equation}
\hat{H}=\hat{H}_{C}+\hat{H}_{E}=\hat{H}_{C}+\hat{H}_{E}^{0}+\hat{V}_{E}=\hat{%
H}_{C}+(\hat{H}_{E}^{0})^{\dagger }+(\hat{V}_{E})^{\dagger }.
\label{QuantH.eq}
\end{equation}
Although there are similarities in that the quantum field Hamiltonian is the
sum of non-commuting cavity and external region contributions, the final
form of our Hamiltonian for the EM\ field differs from that in reference 
\cite{Dutra00}. Here there are no off-diagonal terms so the Hamiltonian is
simpler.

\subsection{\textit{Non-commutation of cavity and external NHM operators}}

The commutation rules for either the cavity NHM or the external region NHM
annihilation and creation operators separately have been described. Since
the field Hamiltonian is the sum of cavity and external region
contributions, each of which depend on their own annihilation and creation
operators, it is also important to consider the commutators between cavity
and external region operators. These can be obtained from those for the
generalised coordinates and momenta and hence depend on certain surface
integrals, as we have seen in equations (\ref{CEcr1.eq} - \ref{CEcr8.eq}).
The commutators are listed in tables 1 and 2. In these tables $a_{nK}=(1/2)\,%
\sqrt{\omega _{n}\,/\,\omega _{K}}$ and $b_{nK}=(1/2)\,\sqrt{\omega
_{K}\,/\,\omega _{n}}$. Another thirty-two commutation rules can be obtained
by taking the adjoints of the results in tables 1 and 2.

\begin{tabular}{||l||l|l|l|l||}
\hline\hline
$\lbrack \hat{L},\hat{R}]$ & $\hat{A}_{K}$ & \multicolumn{1}{||l|}{$\hat{A}%
_{K}^{\#}$} & \multicolumn{1}{||l|}{$\hat{B}_{K}$} & \multicolumn{1}{||l||}{$%
\hat{B}_{K}^{\#}$} \\ \hline\hline
$\hat{A}_{n}$ & $0$ & $a_{nK}\,L_{nK}$ & $-\,a_{nK}\,L_{nK}$ & $0$ \\ 
&  & $+b_{nK}\,\mathcal{L}_{nK}$ & $+b_{nK}\,\mathcal{L}_{nK}$ &  \\ \hline
$\hat{A}_{n}^{\#}$ & $-\,a_{nK}\,K_{nK}^{\ast }$ & $0$ & $0$ & $%
\,a_{nK}\,K_{nK}^{\ast }$ \\ 
& $-b_{nK}\,\mathcal{K}_{nK}^{\ast }$ &  &  & $-b_{nK}\,\mathcal{K}%
_{nK}^{\ast }$ \\ \hline
$\hat{B}_{n}$ & $-\,a_{nK}\,K_{nK}^{\ast }$ & $0$ & $0$ & $%
\,a_{nK}\,K_{nK}^{\ast }$ \\ 
& $+b_{nK}\,\mathcal{K}_{nK}^{\ast }$ &  &  & $+b_{nK}\,\mathcal{K}%
_{nK}^{\ast }$ \\ \hline
$\hat{B}_{n}^{\#}$ & $0$ & $a_{nK}\,L_{nK}$ & $-\,a_{nK}\,L_{nK}$ & $0$ \\ 
&  & $-b_{nK}\,\mathcal{L}_{nK}$ & $-b_{nK}\,\mathcal{L}_{nK}$ &  \\ 
\hline\hline
\end{tabular}

Table 1. Commutators for cavity and external NHM annihilation and creation
operators.

\begin{tabular}{||l||l|l|l|l||}
\hline\hline
$\lbrack \hat{L},\hat{R}]$ & $\hat{A}_{K}^{\dagger }$ & 
\multicolumn{1}{||l|}{$\hat{A}_{K}^{\#\dagger }$} & \multicolumn{1}{||l|}{$%
\hat{B}_{K}^{\dagger }$} & \multicolumn{1}{||l||}{$\hat{B}_{K}^{\#\dagger }$}
\\ \hline\hline
$\hat{A}_{n}$ & $a_{nK}\,J_{nK}$ & $0$ & $0$ & $-\,a_{nK}\,J_{nK}$ \\ 
& $+b_{nK}\,\mathcal{J}_{nK}$ &  &  & $+b_{nK}\,\mathcal{J}_{nK}$ \\ \hline
$\hat{A}_{n}^{\#}$ & $0$ & $-\,a_{nK}\,I_{nK}^{\ast }$ & $%
a_{nK}\,I_{nK}^{\ast }$ & $0$ \\ 
&  & $-b_{nK}\,\mathcal{I}_{nK}^{\ast }$ & $-b_{nK}\,\mathcal{I}_{nK}^{\ast
} $ &  \\ \hline
$\hat{B}_{n}$ & $0$ & $-\,a_{nK}\,I_{nK}^{\ast }$ & $a_{nK}\,I_{nK}^{\ast }$
& $0$ \\ 
&  & $+b_{nK}\,\mathcal{I}_{nK}^{\ast }$ & $+b_{nK}\,\mathcal{I}_{nK}^{\ast
} $ &  \\ \hline
$\hat{B}_{n}^{\#}$ & $a_{nK}\,J_{nK}$ & $0$ & $0$ & $-\,a_{nK}\,J_{nK}$ \\ 
& $-b_{nK}\,\mathcal{J}_{nK}$ &  &  & $-b_{nK}\,\mathcal{J}_{nK}$ \\ 
\hline\hline
\end{tabular}

Table 2. Commutators for cavity and adjoints of external NHM annihilation
and creation operators.

\subsection{\textit{Energy eigenstates for unperturbed Hamiltonians}}

Several possible choices are available for a suitable unperturbed
Hamiltonian for which approximate energy eigenstates can be obtained. One
choice is the cavity region Hamiltonian $\hat{H}_{C}$, another is the
external region approximate Hamiltonian $\hat{H}_{E}^{0}$ . A third choice
is the external region approximate Hamiltonian $(\hat{H}_{E}^{0})^{\dagger }$%
. In the case of $\hat{H}_{C}$ the non-Hermitean quantities $\hat{A}_{n}^{\#}%
\hat{A}_{n},\,\hat{B}_{n}^{\#}\hat{B}_{n}$ behave as a set of commuting
number operators for which we can find simultaneous left and right
eigenstates, even though the individual number operators are non-Hermitean.
These eigenstates are also eigenstates of $\hat{H}_{C}$. Similarly, in the
case of $\hat{H}_{E}^{0}$ the non-Hermitean quantities$\,\hat{A}_{K}^{\#}%
\hat{A}_{K},\,\hat{B}_{K}^{\#}\hat{B}_{K}$\ again behave as a set of
commuting number operators for which we can find simultaneous left and right
eigenstates, even though the individual number operators are non-Hermitean.
These eigenstates are also eigenstates of $\hat{H}_{E}^{0}$. In each case,
the unperturbed Hamiltonian involves the annihilation, creation operator
pairs $(\hat{A}_{n},\hat{A}_{n}^{\#}),(\hat{B}_{n},\hat{B}_{n}^{\#})$ or $(%
\hat{A}_{K},\hat{A}_{K}^{\#}),(\hat{B}_{K},\hat{B}_{K}^{\#})$\ and the true
mode vacuum state $|\,0>\,=|...0_{k}.....0_{k\ast }..>$\ acts approximately
as a vacuum state for the NHM annihilation operators $\hat{A}_{n},\hat{B}%
_{n} $ or $\hat{A}_{K},\hat{B}_{K}$\ , with 
\begin{equation}
\hat{A}_{n}|\,0>=\hat{B}_{n}|\,0>=\hat{A}_{K}|\,0>=\hat{B}_{K}|\,0>=0\ 
\label{VacState.eq}
\end{equation}
(see equations (\ref{approxNHMtrue.eq})). The right simultaneous eigenstates
of the number operators $\hat{A}_{n}^{\#}\hat{A}_{n},\,\hat{B}_{n}^{\#}\hat{B%
}_{n}$ or $\hat{A}_{K}^{\#}\hat{A}_{K},\,\hat{B}_{K}^{\#}\hat{B}_{K}$ can be
constructed in the standard form as:

\begin{eqnarray}
|\{n_{n}\},\{m_{n}\} &>&\,=\prod_{n}\frac{(\hat{A}_{n}^{\#})^{n_{n}}}{\sqrt{%
n_{n}!}}\frac{(\hat{B}_{n}^{\#})^{m_{n}}}{\sqrt{m_{n}!}}|\,\mathbf{0}>,
\label{RtEig1.eq} \\
|\{n_{K}\},\{m_{K}\} &>&\,=\prod_{K}\frac{(\hat{A}_{K}^{\#})^{n_{K}}}{\sqrt{%
n_{K}!}}\frac{(\hat{B}_{K}^{\#})^{m_{K}}}{\sqrt{m_{K}!}}|\,\mathbf{0}>
\label{RtEig2.eq}
\end{eqnarray}
where $n_{n},m_{n},n_{K},m_{K}$\ are all positive integers. These states are
also right eigenstates for the unperturbed Hamiltonians $\hat{H}_{C}$ or $%
\hat{H}_{E}^{0}$, where (with $\{\mathbf{N}_{C}\}\equiv \{n_{n}\},\{m_{n}\}$
and $\{\mathbf{N}_{E}\}=\{n_{K}\},\{m_{K}\}$)

\begin{eqnarray}
\hat{H}_{C}\,|\{\mathbf{N}_{C}\mathbf{\}} &>&\,=E(\{\mathbf{N}_{C}\mathbf{\}}%
)\,\,|\{\mathbf{N}_{C}\mathbf{\}}>,  \label{EnEig1.eq} \\
\hat{H}_{E}^{0}\,|\{\mathbf{N}_{E}\mathbf{\}} &>&\,=E(\{\mathbf{N}_{E}%
\mathbf{\}})\,\,|\{\mathbf{N}_{E}\mathbf{\}}>  \label{EnEig2.eq}
\end{eqnarray}
and the energy is given by:

\begin{eqnarray}
E(\{n_{n}\},\{m_{n}\}) &=&\sum_{n}\,\,\hbar \omega _{n}\,(n_{n}+m_{n}+1)
\label{Energy1.eq} \\
E(\{n_{K}\},\{m_{K}\}) &=&\sum_{K}\,\,\hbar \omega _{K}\,(n_{K}+m_{K}+1)
\label{Energy2.eq}
\end{eqnarray}
These are clearly quantum harmonic oscillator energy expressions. For the
state $|\{n_{n}\},\{m_{n}\}>$\ there are $n_{n},m_{n}$\ photons associated
with the cavity NHMs $U_{n}(R),\,V_{n}(R)^{\ast }$, and for the state $%
|\{n_{K}\},\{m_{K}\}>$\ there are $n_{K},m_{K}$\ photons associated with the
external region NHMs $U_{K}(R),V_{K}(R)^{\ast }$. As the number operators $%
\hat{A}_{n}^{\#}\hat{A}_{n},\,\hat{B}_{n}^{\#}\hat{B}_{n}$ or $\hat{A}%
_{K}^{\#}\hat{A}_{K},\,\hat{B}_{K}^{\#}\hat{B}_{K}$\ are non-Hermitean, the
left simultaneous eigenstates of these operators are different to the right
eigenstates. However, both the left and right eigenstates are also
eigenstates of the unperturbed Hamiltonian $\hat{H}_{C}$ or $\hat{H}%
_{E}^{0}, $\ having the same energy as in equations (\ref{Energy1.eq}, \ref
{Energy2.eq}). The left eigenstates are obtained in the form:

\begin{eqnarray}
(\{n_{n}\},\{m_{n}\}|\, &=&\,<\mathbf{0\,}|\,\prod_{n}\frac{(\hat{A}%
_{n})^{n_{n}}}{\sqrt{n_{n}!}}\frac{(\hat{B}_{n})^{m_{n}}}{\sqrt{m_{n}!}},
\label{LeftEig1.eq} \\
(\{n_{K}\},\{m_{K}\}|\, &=&\,<\mathbf{0\,}|\,\prod_{K}\frac{(\hat{A}%
_{K})^{n_{K}}}{\sqrt{n_{K}!}}\frac{(\hat{B}_{K})^{m_{K}}}{\sqrt{m_{K}!}}
\label{LeftEig2.eq}
\end{eqnarray}
and the eigenstates for $\hat{H}_{C}$ satisfy orthonormality and
completeness relations: 
\begin{eqnarray}
(\{\mathbf{N}_{C}\mathbf{\}\,|\,\{N}_{C}^{/}\mathbf{\}} &>&\mathbf{=\delta (}%
\{\mathbf{N}_{C}\mathbf{\},\{N}_{C}^{/}\mathbf{\})}  \label{Orth.eq} \\
\sum_{\{\mathbf{N}_{C}\mathbf{\}}}|\{\mathbf{N}_{C}\} &>&(\{\mathbf{N}%
_{C}\}|=1  \label{Complete.eq}
\end{eqnarray}
in an obvious notation. The round bracket notation for the left eigenstates
has been introduced to distinguish $(\{\mathbf{N}_{C}\mathbf{\}\,|}$ from
the adjoint of the right eigenstates, $(\mathbf{|\,\{N}_{C}\mathbf{\}>)}%
^{\dagger }=\,<\{\mathbf{N}_{C}\mathbf{\}\,|}$, whose role will be discussed
below. Analogous results apply for the states based on the external
Hamiltonian $\hat{H}_{E}^{0}$. Similar results have been obtained in
reference \cite{Lamprecht99}.

Some simple states will be specified as examples of these expressions. The
zero photon or vacuum state is the same as the true mode vacuum. The one
photon states are: 
\begin{eqnarray}
|1_{_{n}}^{A} &>&=\hat{A}_{n}^{\#}|\,\mathbf{0}>\,\qquad |1_{_{n}}^{B}>=\hat{%
B}_{n}^{\#}|\,\mathbf{0}>\,  \label{OnePhotStat.eq} \\
|1_{_{K}}^{A} &>&=\hat{A}_{K}^{\#}|\,\mathbf{0}>\,\qquad |1_{K}^{B}>=\hat{B}%
_{K}^{\#}|\,\mathbf{0}>.  \notag
\end{eqnarray}

For the cavity Hamiltonian $\hat{H}_{C}$ there are two other choices of
right and left eigenstates, which are the adjoints of the states $%
(\{n_{n}\},\{m_{n}\}|$ and $|\{n_{n}\},\{m_{n}\}>$ and are given by:

\begin{eqnarray}
|\{n_{n}\},\{m_{n}\})\, &=&\prod_{n}\frac{(\hat{A}_{n}^{\dagger })^{n_{n}}}{%
\sqrt{n_{n}!}}\frac{(\hat{B}_{n}^{\dagger })^{m_{n}}}{\sqrt{m_{n}!}}|\,%
\mathbf{0}>  \label{RightEig3.eq} \\
&<&\{n_{n}\},\{m_{n}\}|\,=\,<\mathbf{0\,}|\,\prod_{n}\frac{(\hat{A}_{n}^{\#%
{\Large \dagger }})^{n_{n}}}{\sqrt{n_{n}!}}\frac{(\hat{B}_{n}^{\#{\Large %
\dagger }})^{m_{n}}}{\sqrt{m_{n}!}}.  \label{LeftEig3.eq}
\end{eqnarray}
Both states have energies given by equation (\ref{Energy1.eq}) and the new
right eigenvalue equation is: 
\begin{equation}
\hat{H}_{C}\,|\{\mathbf{N}_{C}\mathbf{\}})\,=E(\{\mathbf{N}_{C}\mathbf{\}}%
)\,\,|\{\mathbf{N}_{C}\mathbf{\}}).  \label{EnEig3.eq}
\end{equation}
The new orthogonality and completeness relationships that apply are: 
\begin{eqnarray}
&<&\{\mathbf{N}_{C}\mathbf{\}\,|\,\{N}_{C}^{/}\mathbf{\})=\delta (}\{\mathbf{%
N}_{C}\mathbf{\},\{N}_{C}^{/}\mathbf{\})}  \label{Orth2.eq} \\
\sum_{\{\mathbf{N}_{C}\mathbf{\}}}|\{\mathbf{N}_{C}\}) &<&\{\mathbf{N}%
_{C}\}|=1.  \label{Complete2.eq}
\end{eqnarray}
These results may be obtained directly by taking the adjoints of previous
equations or by using the alternative form for $\hat{H}_{C}$ in equation (%
\ref{Hc2Dag.eq}), noting that $\hat{A}_{n}^{\dagger }\hat{A}_{n}^{\#\dagger
} $ and $\hat{B}_{n}^{\dagger }\hat{B}_{n}^{\#\dagger }$ are also number
operators and that $\hat{A}_{n}^{\#\dagger }|\,0>=\hat{B}_{n}^{\#\dagger
}|\,0>=0$ using equation (\ref{approxNHMtrue.eq}).

For the approximate external region Hamiltonian the situation is similar.
Here the additional right and left eigenstates of the Hermitean adjoint $(%
\hat{H}_{E}^{0})^{\dagger }$ are $|\{n_{K}\},\{m_{K}\})$ and $%
<\{n_{K}\},\{m_{K}\}|$, which are the adjoints of $(\{n_{K}\},\{m_{K}\}|$
and $|\{n_{K}\},\{m_{K}\}>$ respectively and can be obtained from equations (%
\ref{LeftEig2.eq}, \ref{RtEig2.eq}). Both states have energies given by
equation (\ref{Energy2.eq}) but the right eigenvalue equation now involves $(%
\hat{H}_{E}^{0})^{\dagger }$ and is: 
\begin{equation}
(\hat{H}_{E}^{0})^{\dagger }\,|\{\mathbf{N}_{E}\mathbf{\}})\,=E(\{\mathbf{N}%
_{E}\mathbf{\}})\,\,|\{\mathbf{N}_{E}\mathbf{\}}).  \label{EnEig4.eq}
\end{equation}
Orthogonality and completeness relationships analogous to the cavity case
apply, and the results are obtained similarly.

\section{Atom-field interactions and spontaneous decay}

\label{SectAtom-Field}

\subsection{\textit{Rotating wave approximation and coupling constants}}

In the electric dipole approximation the interaction between an atom and the
field is given by $\hat{V}=\mathbf{\hat{d}\cdot \hat{\Pi}\,/\,\epsilon }_{0}$%
, where $\mathbf{\hat{d}}$ is the atomic dipole operator and the conjugate
momentum field $\mathbf{\hat{\Pi}}$ is evaluated at the atomic position (see
reference \cite{Dalton96}). The dipole operator can be written as the sum of
a component $\mathbf{\hat{d}}^{+}$ associated with upward atomic transitions
and a component $\mathbf{\hat{d}}^{-}$ associated with downward atomic
transitions, where $\mathbf{\hat{d}}^{+}=(\mathbf{\hat{d}}^{-})^{\dagger }$.
The conjugate momentum field can be expressed as the sum of its positive and
negative frequency components, $\mathbf{\hat{\Pi}}^{+},\mathbf{\hat{\Pi}}%
^{-} $ respectively; Thus we have: 
\begin{eqnarray}
\mathbf{\hat{d}} &=&\mathbf{\hat{d}}^{+}+\mathbf{\hat{d}}^{-}
\label{DipSum.eq} \\
\mathbf{\hat{\Pi}} &=&\mathbf{\hat{\Pi}}^{+}+\mathbf{\hat{\Pi}}^{-}.
\label{PiSum.eq}
\end{eqnarray}
Expressions for the positive and negative frequency component for positions $%
\mathbf{R}$ in the cavity are:

\begin{eqnarray}
\mathbf{\hat{\Pi}}^{+} &=&\sum_{n}\frac{1}{i}\sqrt{\hbar \epsilon _{0}\omega
_{n}\,/\,2}(\hat{A}_{n}\,\mathbf{U}_{n}(\mathbf{R)+}\hat{B}_{n}\,\mathbf{V}%
_{n}(\mathbf{R)}^{{\large \ast }}\mathbf{)}  \label{PiPlus.eq} \\
\mathbf{\hat{\Pi}}^{-} &=&-\sum_{n}\frac{1}{i}\sqrt{\hbar \epsilon
_{0}\omega _{n}\,/\,2}(\hat{B}_{n}^{{\large \#}}\,\mathbf{U}_{n}(\mathbf{R)+}%
\hat{A}_{n}^{{\large \#}}\,\mathbf{V}_{n}(\mathbf{R)}^{{\large \ast }}%
\mathbf{)=}(\mathbf{\hat{\Pi}}^{+})^{{\large \dagger }}  \label{PiMinus.eq}
\end{eqnarray}
Note that although $\mathbf{\hat{\Pi}}^{-}\mathbf{=}(\mathbf{\hat{\Pi}}%
^{+})^{\dagger }$ overall, the same is not true on a term by term basis. In
the rotating wave approximation (RWA) the atom-field interaction is given
by: 
\begin{equation}
\hat{V}_{RWA}=(\mathbf{\hat{d}}^{+}\mathbf{\cdot \hat{\Pi}}^{+}+\mathbf{\hat{%
d}}^{-}\mathbf{\cdot \hat{\Pi}}^{-})\mathbf{\,/\,\epsilon }_{0}.
\label{VRWA.def}
\end{equation}
For a two level atom (TLA) with upper state $|e>$, lower state $|g>$, the
dipole operator components are: $\mathbf{\hat{d}}^{\pm }=\mathbf{d}\,\hat{%
\sigma}_{\pm }$ , where $\mathbf{d}$ is the dipole matrix element $<e\,|\,%
\mathbf{\hat{d}\,}|\,g>$ and $\hat{\sigma}_{+}=|e><g|,\,\hat{\sigma}%
_{-}=|g><e|$ are the upward, downward transition operators. The atomic
transition energy is $\hbar \omega _{0}$. For the TLA located in the cavity
at position $\mathbf{R}$, the RWA atom-field interaction can be expressed as:

\begin{equation}
\hat{V}_{RWA}=\hat{\sigma}_{+}\,[\sum_{n}\,\hbar (g_{n}^{A}\,\hat{A}%
_{n}+g_{n}^{B}\,\hat{B}_{n})]-\hat{\sigma}_{-}\,[\sum_{n}\,\hbar (g_{n}^{A}\,%
\hat{B}_{n}^{{\Large \#}}+g_{n}^{B}\,\hat{A}_{n}^{{\Large \#}})],
\label{VRWA.eq}
\end{equation}
where the coupling constants are defined by: 
\begin{eqnarray}
g_{n}^{A} &=&-i\sqrt{\omega _{n}\,/\,2\hbar \epsilon _{0}}(\mathbf{d\cdot U}%
_{n}\mathbf{(R))}  \label{gAcoup.eq} \\
g_{n}^{B} &=&-i\sqrt{\omega _{n}\,/\,2\hbar \epsilon _{0}}(\mathbf{d\cdot V}%
_{n}\mathbf{(R)}^{\ast }\mathbf{)}.  \label{gbcoup.eq}
\end{eqnarray}
These coupling constants are also referred to as one-photon Rabi
frequencies. Here we note one of the unusual features of the NHM theory -
whereas an upward atomic transition accompanied by the \textit{absorption}
of a $\mathbf{U}_{n}(\mathbf{R)}$ photon will be associated with the
coupling constant $g_{n}^{A}$, a downward atomic transition accompanied by
the \textit{emission} of a $\mathbf{U}_{n}(\mathbf{R)}$ photon will be
associated with the \textit{opposite} coupling constant $g_{n}^{B}$. The
situation involving the absorption or emission of a $\mathbf{V}_{n}(\mathbf{%
R)}^{\ast }$photon is similar, with the \textit{opposite} coupling constant $%
g_{n}^{A}$ again being associated with emission. Thus absorption and
emission involve different coupling constants, a peculiarity ultimately
responsible for the Petermann factor.

\subsection{\textit{Decay of excited atom}}

The spontaneous emission problem for the TLA located in the cavity will be
treated via a simple approach using the essential states approximation and
the RWA. In addition, the effect of the external region term $\hat{H}_{E}$
will be neglected in the first instance, where we just focus on the initial
emission of photons into the cavity NHM. The external region term $\hat{H}%
_{E}$ comes into play when the loss of cavity NHM photons and the creation
of external NHM photons is considered - that is, cavity loss processes. As
this implies, we are assuming that the cavity loss rate is slow compared to
the spontaneous emission rate from the atom into the cavity. Thus we will
take as our Hamiltonian: 
\begin{equation}
\hat{H}_{T}=\hat{H}_{A}+\hat{H}_{C}+\hat{V}_{RWA},  \label{Htot.eq}
\end{equation}
where $\hat{H}_{A}$ is the atomic Hamiltonian and ignoring the zero point
energy terms as usual. Hence our essential states will just include product
states of the form: $|e>|\mathbf{0>,|}g>|1_{n}^{A}>$ and $\mathbf{|}%
g>|1_{n}^{B}>$. The initial state is given by: 
\begin{equation}
|\Psi (0)>=|e>|\mathbf{0>,}  \label{Psi0.eq}
\end{equation}
and the time dependent Schrodinger state vector is written as:

\begin{eqnarray}
|\Psi (t) &>&=C_{e}(t)\,\exp (-i\omega _{0}t/2)\,|e>|\mathbf{0>}  \notag \\
+\,\sum_{n}\,C_{n}^{A}(t)\,\exp (-i[\omega _{n}-\omega _{0}/2]t)\,\mathbf{|}%
g &>&|1_{n}^{A}> \\
+\,\sum_{n}\,C_{n}^{B}(t)\,\exp (-i[\omega _{n}-\omega _{0}/2]t)\,\mathbf{|}%
g &>&|1_{n}^{B}>,
\end{eqnarray}
where $C_{e}(t),C_{n}^{A}(t)$ and $C_{n}^{B}(t)$ are amplitudes in the
interaction picture. Using the time dependent Schrodinger equation we obtain
coupled equations for the amplitudes:

\begin{eqnarray}
i\frac{dC_{e}(t)}{dt} &=&-\,\sum_{n}\,\{g_{n}^{A}\,\exp (i\delta
_{n}t)\,C_{n}^{A}(t)+g_{n}^{B}\,\exp (i\delta _{n}t)\,C_{n}^{B}(t)\} \\
i\frac{dC_{n}^{A}(t)}{dt} &=&g_{n}^{B}\,\exp (-i\delta _{n}t)\,C_{e}(t) \\
i\frac{dC_{n}^{B}(t)}{dt} &=&g_{n}^{A}\,\exp (-i\delta _{n}t)\,C_{e}(t),
\end{eqnarray}
where the detuning is $\delta _{n}=\omega _{0}-\omega _{n}$. Note the
opposite coupling constants in the last two equations. The one photon
amplitudes $C_{n}^{A}(t),C_{n}^{B}(t)$ can be formally eliminated to give a
single integro-differential equation for the excited state amplitude $%
C_{e}(t)$ of the form:

\begin{equation}
\frac{dC_{e}(t)}{dt}=\int_{0}^{t}\,d\tau \,K(\tau )\,C_{e}(t-\tau ),
\label{IntDE.eq}
\end{equation}
where the kernel $K(\tau )$ is defined by: 
\begin{equation}
K(\tau )=\sum_{n}2\,g_{n}^{A}\,g_{n}^{B}\,\exp (i\delta _{n}\tau ).
\label{Ktau.def}
\end{equation}
In view of the convolution integral form, the equation for $C_{e}(t)$ can be
solved in the general case via Laplace transform methods in terms of the
Laplace transform of the kernel: 
\begin{equation}
\tilde{K}(s)=i\sum_{n}2\,g_{n}^{A}\,g_{n}^{B}/(\delta _{n}+is).
\label{Ks.eq}
\end{equation}

Recalling that our generic index $n$ specfies the angular frequency $\omega
_{n}$, the polarization $\mathbf{\alpha }_{n}$ and the transverse mode index 
$\theta _{n}$, we may convert the sum over $n$ in equation (\ref{Ktau.def})
to an integral over $\omega _{n}$ times a mode density $\rho (\omega _{n})$,
together with sums over $\mathbf{\alpha }_{n}$ and $\theta _{n}$. Then,
assuming that the coupling constants $g_{n}^{A}\,,g_{n}^{B}$ are slowly
varying with frequency $\omega _{n}$, we see that the kernel $K(\tau )$
decreases to zero over a correlation time $\tau _{c}$ of order the inverse
bandwidth of the coupling constants. Further, assuming this time scale is
short compared to the decay time $T_{e}$ of the excited state amplitude, we
can then make the Markoff approximation. In this case:

\begin{equation}
\frac{dC_{e}(t)}{dt}\simeq \{\int_{0}^{\infty }\,d\tau \,K(\tau
)\}\,C_{e}(t),  \label{CeMark.eq}
\end{equation}
showing that the probability of the atom being excited, $%
P_{e}(t)=|C_{e}(t)|^{2}$ decays exponentially to zero with a rate $\Gamma
_{e}$ given by: 
\begin{equation}
\Gamma _{e}=-2\,\func{Re}\tilde{K}(\epsilon ),  \label{Gamma.def}
\end{equation}
where $\epsilon $ is small.

To evaluate $\Gamma _{e}$, equation (\ref{UV.eq}) can be used to interrelate
the coupling constants in equation (\ref{Ks.eq}), giving 
\begin{equation}
g_{n}^{B}=-\sum_{m}\,\mathbf{D}_{nm}\,(g_{m}^{A})^{\ast },  \label{gBgA.eq}
\end{equation}
which is then substituted into equation (\ref{Ks.eq}). Now the
transformation matrix $\mathbf{D}_{nm}$ is zero unless the frequency and
polarizations are the same ($\omega _{n}=\omega _{m},\mathbf{\alpha }_{n}=%
\mathbf{\alpha }_{m}$), so after replacing the sums over $n$ and $m$ by an
integral over $\omega _{n}$ involving NHM density $\rho (\omega _{n})$ and
by sums over $\mathbf{\alpha }_{n}$ and the transverse mode indices $\theta
_{n},\theta _{m}$, we then obtain for $\tilde{K}(\epsilon )$ the expression:

\begin{eqnarray}
\tilde{K}(\epsilon ) &=&i\,\int \,d\omega _{n}\,\rho (\omega
_{n})\,2\,\sum_{\,\mathbf{\alpha }_{n}}\,\sum_{\,\theta
_{n}}\,\sum_{\,\theta _{m}}\,(g_{n}^{A})\,\mathbf{D}_{nm}\,(g_{m}^{A})^{\ast
}\,\frac{P}{(\omega _{n}-\omega _{0})}  \notag \\
&&-\pi \int \,d\omega _{n}\,\rho (\omega _{n})\,2\,\,\sum_{\,\alpha
_{n}}\,\sum_{\,\theta _{n}}\,\sum_{\,\theta _{m}}\,(g_{n}^{A})\,\mathbf{D}%
_{nm}\,(g_{m}^{A})^{\ast }\,\delta (\omega _{n}-\omega _{0}).  \label{Ke.eq}
\end{eqnarray}
From the Hermiteancy properties of $\mathbf{D}_{nm}$ it is not difficult to
see that the double sum $\sum_{\,\theta _{n}}\,\sum_{\,\theta
_{m}}\,(g_{n}^{A})\,\mathbf{D}_{nm}\,(g_{m}^{A})^{\ast }$ is real, so that
in general we have for the decay rate: 
\begin{eqnarray}
\Gamma _{e} &=&4\pi \rho (\omega _{0})\sum_{\mathbf{\,\alpha }%
_{n}}\,\sum_{\,\theta _{n}}\,|g_{n}^{A}(\omega _{0})|^{2}\,K_{n}(\omega
_{0})\,  \notag \\
&&+4\pi \rho (\omega _{0})\sum_{\mathbf{\,\alpha }_{n}}\,\sum_{\,\theta
_{n}\neq \theta _{m}}\,\sum_{\,\theta _{m}\neq \theta
_{n}}\,g_{n}^{A}(\omega _{0})\,\mathbf{D}_{nm}(\omega
_{0})\,g_{m}^{A}(\omega _{0})^{\ast },  \label{GammaKGen.eq}
\end{eqnarray}
where we have substituted the Petermann factors $K_{n}(\omega _{0})$ for the
diagonal matrix elements $\mathbf{D}_{nn}(\omega _{0})$. Note that all
(diagonal and off-diagonal) transformation matrix elements $\mathbf{D}_{nm}$
with $\mathbf{\alpha }_{n}=\,\mathbf{\alpha }_{m}$ are independent of $%
\mathbf{\alpha }_{n}$. In general then, the decay rate $\Gamma _{e}$ will
involve the cavity NHM coupling constants $g_{n}^{A}$ and the transformation
matrix elements $\mathbf{D}_{nm}$ evaluated at the atomic transition
frequency $\omega _{0}$. The decay rate is the sum of diagonal terms
involving the Petermann factors $K_{n}(\omega _{0})=\mathbf{D}_{nn}(\omega
_{0})$ and off-diagonal terms involving the general $\mathbf{D}_{nm}(\omega
_{0})$ with $\,\theta _{n}\neq \theta _{m}$. However, if the transformation
matrix is such that the sums over $\theta _{n},\theta _{m}$ are dominated by
the contribution coming from a \textit{single} (real) diagonal element $%
\mathbf{D}_{nn}$ with transverse mode index $\theta _{n}$, then the other
terms can be ignored, leading to the following simple result for the decay
rate:

\begin{eqnarray}
\Gamma _{e} &=&4\pi \rho (\omega _{0})\sum_{\alpha _{n}}\,|g_{n}^{A}(\omega
_{0},\theta _{n})|^{2}\,\mathbf{D}_{nn}(\omega _{0},\theta _{n})  \notag \\
&=&K_{n}\,\Gamma _{e}^{F},  \label{GammaK.eq}
\end{eqnarray}
where 
\begin{equation}
\Gamma _{e}^{F}=4\pi \rho (\omega _{0})\sum_{\alpha _{n}}\,|g_{n}^{A}(\omega
_{0},\theta _{n})|^{2}  \label{GammaFree.eq}
\end{equation}
is the decay rate expected for a normal cavity situation. Thus we see for
this special case where a single NHM dominates, the decay rate has been
enhanced by the Petermann factor $K_{n}$.

\section{Conclusion}

\label{SectConc}

The present paper is a fully quantum treatment of the field for unstable
optical systems. A standard canonical quantization proceedure is used, based
on expanding the vector potential in the unstable cavity region and in the
external region via non-Hermitean (Fox-Li) modes and their Hermitean adjoint
modes, all defined via the optical system. Three dimensional systems are
treated, using the paraxial \cite{Lax75} and monochromaticity approximations 
\cite{Lamprecht99} for the cavity non-Hermitean modes. Both right and left
travelling modes are included. The results are similar to \cite{Lamprecht99}%
, \cite{Dutra00} but differ in detail. The field is equivalent to a set of
quantum harmonic oscillators (QHO), associated now with non-Hermitean modes
rather than true modes, and thus confirming the validity of the photon model
for the case of unstable optical systems. The annihilation, creation
operator pairs for each QHO are not Hermitean adjoints. A doubling of the
number of annihilation, creation operator pairs occurs compared to \cite
{Lamprecht99}, \cite{Dutra00}, showing that the total number of true mode
QHO's equals the total number of QHO's associated either with the cavity or
the external region non-Hermitean modes. The final form of our quantum
Hamiltonian for the EM field is the sum of non-commuting cavity and external
field contributions. To a good approximation, both the cavity and external
field Hamiltonians can be expressed as a sum of independent QHO Hamiltonians
for each non-Hermitean mode, but the external field Hamiltonian also
includes a coupling term responsible for external non-Hermitean mode photon
exchange processes. The two independent QHO Hamiltonians are alternative
choices for an unperturbed Hamiltonian. Left and right eigenstates for each
of these unperturbed Hamiltonians are obtained, and the energy is given by
the usual QHO result. Certain cavity non-Hermitean mode annihilation,
creation operators do not commute with others for the external region
non-Hermitean modes, leading to cavity energy gain and loss processes. A
simple description of the cavity-external region light coupling in terms of
surface integrals involving products of cavity and external non-Hermitean
mode functions is found.

Atom-field interactions are treated in the electric dipole and rotating wave
approximations. Atomic transitions leading to cavity non-Hermitean mode
photon absorption are associated with a different coupling constant to that
for atomic transitions leading to photon emission. This feature is directly
related to the treatment being based on non-Hermitean mode functions and
leads to enhanced emission rates. Using the essential states approach, the
spontaneous decay of a two level atom located in the cavity is treated. The
external region term in the field Hamiltonian is neglected assuming that
atomic decay into the cavity is much faster than cavity decay. Coupled
equations for the amplitudes of atom-field states involving the excited atom
and no photons or the ground state atom with one photon in a cavity
non-Hermitean mode are obtained. Markovian decay occurs under certain
conditions, the decay rate being enhanced by the Petermann factor \cite
{Petermann79} in special cases when a single NHM dominates.

\section{Acknowledgements}

This work was supported by the Engineering and Physical Sciences Research
Council (U.K.). Helpful discussions with Professor M. Babiker, Professor S.
M. Barnett, Professor P. D. Drummond, Dr. S. Dutra, Dr. B. M. Garraway,
Professor P. L. Knight, Dr. C. Lamprecht, Dr. P. W. Milonni, Professor G. C.
H. New, Dr. G. Nienhuis, Professor H. Ritsch and Professor S. Stenholm are
gratefully acknowledged.

\section{Appendix A: Commutation rules for cavity and external region
operators}

\label{Appendix A}

\subsection{\textit{Commutation rules for \^{D} and \^{B} field operators}}

To derive the commutation rules between cavity NHM generalised coordinates
or momenta and external region NHM generalised momenta or coordinates we use
the general rules for the magnetic field and electric displacement operators:

\begin{eqnarray}
\lbrack \mathbf{\hat{B}}^{\alpha }\mathbf{(R),\hat{D}}^{\beta }\mathbf{(R}%
^{/}\mathbf{)]} &=&\mathbf{\,}i\hbar \,\varepsilon _{\alpha \beta \gamma
}\,\partial /\partial X_{\gamma }\,\delta (\mathbf{R-R}^{/})
\label{ApCRBD.eq} \\
\lbrack \mathbf{\hat{D}}^{\alpha }\mathbf{(R),\hat{B}}^{\beta }\mathbf{(R}%
^{/}\mathbf{)]} &=&\mathbf{-\,}i\hbar \,\varepsilon _{\alpha \beta \gamma
}\,\partial /\partial X_{\gamma }\,\delta (\mathbf{R-R}^{/}),
\label{ApCRDB.eq}
\end{eqnarray}
where $\mathbf{R}$ is in the cavity region and $\mathbf{R}^{/}$ is in the
external region. In the cavity region we will use the lowest order terms in
the small parameter $f$ for both $\mathbf{\hat{B}(R)}$ and $\mathbf{\hat{D}%
(R)}$ namely: 
\begin{eqnarray}
\mathbf{\hat{B}(R)} &=&\sum_{n}(\,ik_{n}\,\hat{Q}_{n}\,\mathbf{\hat{z}\times
U}_{n}(\mathbf{R})-ik_{n}\,\hat{R}_{n}^{\dagger }\,\mathbf{\hat{z}\times V}%
_{n}^{\ast }(\mathbf{R}))  \label{ApBCav.eq} \\
\mathbf{\hat{D}(R)} &=&\mathbf{-\,}\sum_{n}(\,\hat{S}_{n}\,\mathbf{U}_{n}(%
\mathbf{R})+\,\hat{P}_{n}^{\dagger }\,\mathbf{V}_{n}^{\ast }(\mathbf{R})),
\label{ApDCav.eq}
\end{eqnarray}
whilst in the external region we will write: 
\begin{eqnarray}
\mathbf{\hat{D}(\mathbf{R}^{/})} &=&\mathbf{-\,}\sum_{K}(\,\hat{S}_{K}\,%
\mathbf{U}_{K}(\mathbf{R}^{/})+\,\hat{P}_{K}^{\dagger }\,\mathbf{V}%
_{K}^{\ast }(\mathbf{R}^{/}))  \label{ApDExt.eq} \\
\mathbf{\hat{B}(\mathbf{R}^{/})} &=&\sum_{K}(\,\hat{Q}_{K}\,\nabla \mathbf{%
\times U}_{K}(\mathbf{R}^{/})+\,\hat{R}_{K}^{\dagger }\,\nabla \mathbf{%
\times V}_{K}^{\ast }(\mathbf{R}^{/})).  \label{ApBExt.eq}
\end{eqnarray}
For simplicity the cavity will be taken to be the region between the planes $%
z=z_{0}$ and $z=z_{b}+\epsilon $ and the external region to be the region
between $z=z_{b}-\epsilon $ and $z=\infty $. The quantity $\epsilon $ is a
small distance to allow for the cavity and external regions to overlap
slightly, and will be taken to zero afterwards. The plane $z=z_{b}$ is the
boundary between the cavity and external regions, which are shown in figure
2.

\subsection{\textit{Cases of cavity NHM coordinates and external NHM momenta}%
}

The commutation rules between cavity NHM coordinates and external region
momenta are found by substituting for $\mathbf{\hat{B}(R)}$ and $\mathbf{%
\hat{D}(\mathbf{R}^{/})}$ from equations (\ref{ApBCav.eq}, \ref{ApDExt.eq})
into equation (\ref{ApCRBD.eq}), which gives:

\begin{eqnarray}
\mathbf{\,}i\hbar \,\sum_{\gamma }\varepsilon _{\alpha \beta \gamma
}\,\partial /\partial X_{\gamma }\,\delta (\mathbf{R-R}^{/})
&=&-\,\sum_{\gamma m(\gamma )}\sum_{L}\{\,ik_{m}\,\varepsilon _{\alpha
z\gamma }U_{m}(\mathbf{R})\,U_{L}^{\beta }(\mathbf{R}^{/})\,[\hat{Q}_{m},\,%
\hat{S}_{L}]\,  \notag \\
&&\qquad \qquad +\,ik_{m}\,\varepsilon _{\alpha z\gamma }U_{m}(\mathbf{R}%
)\,V_{L}^{\beta \ast }(\mathbf{R}^{/})\,[\hat{Q}_{m},\,\hat{P}_{L}^{\dagger
}]  \notag \\
&&\qquad \qquad -\,ik_{m}\,\varepsilon _{\alpha z\gamma }V_{m}^{\ast }(%
\mathbf{R})\,U_{L}^{\beta }(\mathbf{R}^{/})\,[\hat{R}_{m}^{\dagger },\,\hat{S%
}_{L}]\,  \notag \\
&&\qquad \qquad -\,ik_{m}\,\varepsilon _{\alpha z\gamma }V_{m}^{\ast }(%
\mathbf{R})\,V_{L}^{\beta \ast }(\mathbf{R}^{/})\,[\hat{R}_{m}^{\dagger },\,%
\hat{P}_{L}^{\dagger }]\mathbf{\,}\}.  \notag \\
&&  \label{Ap1.eq}
\end{eqnarray}
On the right side the sum over $\gamma $ only involves $x,y$ and that over $%
m $ only those with $\mathbf{\hat{\alpha}}_{m}=\mathbf{\hat{\gamma}}$. For
such terms the components of $\mathbf{U}_{m}$ and $\mathbf{V}_{m}^{\ast }$
are $U_{m}$ and $V_{m}^{\ast }$.

\subsubsection{\textit{Case of }$\,[\hat{Q}_{n},\,\hat{P}_{K}^{\dagger }]$}

To obtain the commutation rule for $\,[\hat{Q}_{n},\,\hat{P}_{K}^{\dagger }]$
we multiply each side of equation (\ref{Ap1.eq}) by $V_{n}^{\ast }(\mathbf{R}%
)\,\,U_{K}^{\beta }(\mathbf{R}^{/})$, integrate $\mathbf{R,R}^{/}$ over the
cavity and external regions respectively, then sum over $\beta $. Using the
biorthogonality results in equations (\ref{Biorth.eq}, \ref{Biorth2.eq}) and
neglecting integrals where the complex conjugation occurs either zero or two
times we find that: 
\begin{equation}
\mathbf{\,}i\hbar \,\sum_{\beta \gamma }\varepsilon _{\alpha \beta \gamma
}\,\int_{C}d^{3}\mathbf{R\,}\int_{E}d^{3}\mathbf{R}^{/}\,V_{n}^{\ast }(%
\mathbf{R})\,\,U_{K}^{\beta }(\mathbf{R}^{/})\,\partial /\partial X_{\gamma
}\,\delta (\mathbf{R-R}^{/})=-\sum_{\gamma =\alpha
_{n}}\,ik_{n}\,\varepsilon _{\alpha z\gamma }\,[\hat{Q}_{n},\,\hat{P}%
_{K}^{\dagger }].  \label{Ap3.eq}
\end{equation}
There are only two non-zero possibilities $\mathbf{\hat{\alpha}}=\mathbf{%
\hat{x}}$ or $\mathbf{\hat{\alpha}}=\mathbf{\hat{y}}$. These give the two
equations: 
\begin{eqnarray}
\,ik_{n}\,[\hat{Q}_{n},\,\hat{P}_{K}^{\dagger }]. &=&\mathbf{\,}i\hbar
\,\int_{C}d^{3}\mathbf{R\,}\int_{E}d^{3}\mathbf{R}^{/}\,V_{n}^{\ast }(%
\mathbf{R})\,\,U_{K}^{y}(\mathbf{R}^{/})\,\partial /\partial z\,\delta (%
\mathbf{R-R}^{/})  \notag \\
&&-\mathbf{\,}i\hbar \,\int_{C}d^{3}\mathbf{R\,}\int_{E}d^{3}\mathbf{R}%
^{/}\,V_{n}^{\ast }(\mathbf{R})\,\,U_{K}^{z}(\mathbf{R}^{/})\,\partial
/\partial y\,\delta (\mathbf{R-R}^{/})  \notag \\
&&  \label{Ap4.eq}
\end{eqnarray}
with $\mathbf{\hat{\alpha}}_{n}=\mathbf{\hat{y}}$ and 
\begin{eqnarray}
-ik_{n}\,[\hat{Q}_{n},\,\hat{P}_{K}^{\dagger }]. &=&\mathbf{\,}i\hbar
\,\int_{C}d^{3}\mathbf{R\,}\int_{E}d^{3}\mathbf{R}^{/}\,V_{n}^{\ast }(%
\mathbf{R})\,\,U_{K}^{z}(\mathbf{R}^{/})\,\partial /\partial x\,\delta (%
\mathbf{R-R}^{/})  \notag \\
&&-\mathbf{\,}i\hbar \,\int_{C}d^{3}\mathbf{R\,}\int_{E}d^{3}\mathbf{R}%
^{/}\,V_{n}^{\ast }(\mathbf{R})\,\,U_{K}^{x}(\mathbf{R}^{/})\,\partial
/\partial z\,\delta (\mathbf{R-R}^{/})  \notag \\
&&  \label{Ap5.eq}
\end{eqnarray}
with $\mathbf{\hat{\alpha}}_{n}=\mathbf{\hat{x}}$.

Considering the double integral on the right side of equation (\ref{Ap5.eq})
involving $\partial /\partial x$, this is of the form: 
\begin{eqnarray}
I_{x} &=&\int_{C}dy\,dz\,\int_{E}d^{3}\mathbf{R}^{/}\,\delta (y\mathbf{-}%
y^{/})\,\delta (z\mathbf{-}z^{/})\,U_{K}^{z}(\mathbf{R}^{/})\,  \notag \\
&&\qquad .\int_{C}dx\,V_{n}^{\ast }(\mathbf{R})\,\partial /\partial
x\,\delta (x\mathbf{-}x^{/})  \notag \\
&=&\int_{C}dy\,dz\,\int_{E}d^{3}\mathbf{R}^{/}\,\delta (y\mathbf{-}%
y^{/})\,\delta (z\mathbf{-}z^{/})\,U_{K}^{z}(\mathbf{R}^{/})\,  \notag \\
&&\qquad .\int_{C}dx\,\{\,\partial /\partial x\,[\delta (x\mathbf{-}%
x^{/})V_{n}^{\ast }(\mathbf{R})]-\delta (x\mathbf{-}x^{/})\,\,\partial
/\partial x\,[V_{n}^{\ast }(\mathbf{R})]\},  \notag \\
&&
\end{eqnarray}
using partial integration. For the first term carrying out the $x$
integration leads to a result dependent on $V_{n}^{\ast }(\mathbf{R})$ for $%
x=-\infty $ and $x=+\infty $. As the NHM functions eventually go to zero for
large $x$, this contribution can be ignored. For the remaining term we now
have: 
\begin{eqnarray}
I_{x} &=&-\int_{C}d^{3}\mathbf{R\,}\int_{E}d^{3}\mathbf{R}^{/}\,\delta (%
\mathbf{R-R}^{/})\,\,U_{K}^{z}(\mathbf{R}^{/})\,\,\partial /\partial
x\,[V_{n}^{\ast }(\mathbf{R})]  \notag \\
&=&-\int_{C}d^{2}\mathbf{s\,}\int_{E}d^{2}\mathbf{s}^{/}\,\delta (\mathbf{s-s%
}^{/})\,\,U_{K}^{z}(\mathbf{s}^{/},z_{b})\,\,\partial /\partial
x\,[V_{n}^{\ast }(\mathbf{s,}z_{b})]  \notag \\
&&\qquad \qquad .\int_{z_{0}}^{z_{b}+\epsilon }\,dz\int_{z_{b}-\epsilon
}^{\infty }dz^{/}\,\delta (z\mathbf{-}z^{/}),
\end{eqnarray}
where in view of the $\delta (\mathbf{R-R}^{/})$ factor, the ordinary
functions can be expressed in terms of their values on the boundary, where $%
z=z_{b}$. The integrals over $z$ and $z^{/}$ give a result proportional to
the small quantity $\epsilon $, and thus the integral $I_{x}$ is zero.
Similar considerations show that the term involving $\partial /\partial y$
on the right side of equation (\ref{Ap4.eq}) is also zero. The two equations
(\ref{Ap4.eq}, \ref{Ap5.eq}) are now both the same as: 
\begin{eqnarray}
\,[\hat{Q}_{n},\,\hat{P}_{K}^{\dagger }]. &=&(\mathbf{\,}\hbar
\,/\,k_{n})\int_{C}d^{3}\mathbf{R\,}\int_{E}d^{3}\mathbf{R}^{/}\,\mathbf{V}%
_{n}^{\ast }(\mathbf{R})\cdot \,\,\mathbf{U}_{K}(\mathbf{R}^{/})\,\partial
/\partial z\,\delta (\mathbf{R-R}^{/})  \notag \\
&&  \label{App6.eq}
\end{eqnarray}
since for the case where $\mathbf{\hat{\alpha}}_{n}=\mathbf{\hat{x}}$ the
scalar product picks out the component $U_{K}^{x}(\mathbf{R}^{/})$ and for
the case $\mathbf{\hat{\alpha}}_{n}=\mathbf{\hat{y}}$ the scalar product
picks out the component $U_{K}^{y}(\mathbf{R}^{/})$.

Considering the double integral on the right side of equation (\ref{App6.eq}%
) involving $\partial /\partial z$, this is of the form: 
\begin{eqnarray}
I_{z} &=&-\int_{C}d^{3}\mathbf{R\,}\int_{E}d^{3}\mathbf{R}^{/}\,\mathbf{V}%
_{n}^{\ast }(\mathbf{R})\cdot \,\,\mathbf{U}_{K}(\mathbf{R}^{/})\,\,\partial
/\partial z\,[\delta (\mathbf{R-R}^{/})]  \notag \\
&=&\int_{C}dx\,dy\,\int_{E}d^{3}\mathbf{R}^{/}\,\delta (x\mathbf{-}%
x^{/})\,\delta (y\mathbf{-}y^{/})\,\mathbf{U}_{K}(\mathbf{R}^{/})\,  \notag
\\
&&\cdot \int_{C}dz\,\{\,\partial /\partial z\,[\delta (z\mathbf{-}z^{/})%
\mathbf{V}_{n}^{\ast }(\mathbf{R})]-\delta (z\mathbf{-}z^{/})\,\,\partial
/\partial z\,[\mathbf{V}_{n}^{\ast }(\mathbf{R})]\}  \notag \\
&=&\int_{C}d^{2}\mathbf{s}\,\int_{E}d^{2}\mathbf{s}^{/}\,\delta (\mathbf{s-s}%
^{/})\,\int_{z_{b}-\epsilon }^{\infty }dz^{/}\,\,\mathbf{U}_{K}(\mathbf{s}%
^{/},z^{/})  \notag \\
&&\cdot \,\int_{z_{0}}^{z_{b}+\epsilon }dz\,\{\,\partial /\partial
z\,[\delta (z\mathbf{-}z^{/})\mathbf{V}_{n}^{\ast }(\mathbf{s,}z)]-\delta (z%
\mathbf{-}z^{/})\,\,\partial /\partial z\,[\mathbf{V}_{n}^{\ast }(\mathbf{s,}%
z)]\}  \notag \\
&&
\end{eqnarray}
using partial integration and then specifying the limits of integration over 
$z$ and $z^{/}$. The integration over $z$ in the first term is carried out
and unlike the similar integrations over $x$ or $y$ that applied to the $%
I_{x}$ or $I_{y}$ cases we find that a non-zero contribution results from
the $z=z_{b}+\epsilon $ limit. The second term is similar to the other term
in the $I_{x}$ or $I_{y}$ cases and overall we now have: 
\begin{eqnarray}
I_{z} &=&\int_{C}d^{2}\mathbf{s}\,\int_{E}d^{2}\mathbf{s}^{/}\,\delta (%
\mathbf{s-s}^{/})\,\int_{z_{b}-\epsilon }^{\infty }dz^{/}\,\,\mathbf{U}_{K}(%
\mathbf{s}^{/},z^{/})  \notag \\
&&\qquad \cdot \{\delta (z_{b}+\epsilon \mathbf{-}z^{/})\mathbf{V}_{n}^{\ast
}(\mathbf{s,}z_{b}+\epsilon )\}  \notag \\
&&-\int_{C}d^{2}\mathbf{s}\,\int_{E}d^{2}\mathbf{s}^{/}\,\delta (\mathbf{s-s}%
^{/})\,\int_{z_{b}-\epsilon }^{\infty }dz^{/}\,\,\mathbf{U}_{K}(\mathbf{s}%
^{/},z^{/})  \notag \\
&&\qquad \cdot \,\int_{z_{0}}^{z_{b}+\epsilon }dz\,\{\,\delta (z\mathbf{-}%
z^{/})\,\,\partial /\partial z\,[\mathbf{V}_{n}^{\ast }(\mathbf{s,}z)]\} 
\notag \\
&=&\int_{C}d^{2}\mathbf{s}\,\int_{E}d^{2}\mathbf{s}^{/}\,\delta (\mathbf{s-s}%
^{/})\,\mathbf{V}_{n}^{\ast }(\mathbf{s,}z_{b})  \notag \\
&&\qquad \cdot \mathbf{U}_{K}(\mathbf{s}^{/},z_{b})\int_{z_{b}-\epsilon
}^{\infty }dz^{/}\,\,\delta (z_{b}+\epsilon \mathbf{-}z^{/})  \notag \\
&&-\int_{C}d^{2}\mathbf{s}\,\int_{E}d^{2}\mathbf{s}^{/}\,\delta (\mathbf{s-s}%
^{/})\,\mathbf{U}_{K}(\mathbf{s}^{/},z_{b})\cdot \partial /\partial z\,[%
\mathbf{V}_{n}^{\ast }(\mathbf{s,}z)]\,_{z_{b}}  \notag \\
&&\qquad \qquad \int_{z_{b}-\epsilon }^{\infty
}dz^{/}\,\,\,\int_{z_{0}}^{z_{b}+\epsilon }dz\,\,\delta (z\mathbf{-}z^{/})\,.
\label{Ap9.eq}
\end{eqnarray}
In equation (\ref{Ap9.eq}) the first term is non-zero, since the $z^{/}\,$%
integral equals one and we are left with a surface integral over the
cavity-external region boundary when the $\delta (\mathbf{s-s}^{/})\,$factor
is integrated out. The second term in equation (\ref{Ap9.eq}) is zero, the
integrals over $z$ and $z^{/}$ giving a result proportional to the small
quantity $\epsilon $, and as in the $I_{x}$ case the overall integral is
zero. Hence we have: 
\begin{equation}
I_{z}=\int_{S}d^{2}\mathbf{s}\,\,\mathbf{V}_{n}^{\ast }(\mathbf{s,}%
z_{b})\cdot \mathbf{U}_{K}(\mathbf{s},z_{b})  \label{Ap10.eq}
\end{equation}
and thus 
\begin{equation}
\lbrack \hat{Q}_{n},\,\hat{P}_{K}^{\dagger }].=(\mathbf{\,}\hbar
\,/\,k_{n})\,\int_{S}d^{2}\mathbf{s}\,\,\mathbf{V}_{n}^{\ast }(\mathbf{s,}%
z_{b})\cdot \mathbf{U}_{K}(\mathbf{s},z_{b}).  \label{Ap11.eq}
\end{equation}

\subsubsection{\textit{Case of }$\,[\hat{R}_{n},\,\hat{S}_{K}^{\dagger }]$}

To obtain the commutation rule for $\,[\hat{R}_{n},\,\hat{S}_{K}^{\dagger }]$%
, we first obtain the result for $\,[\hat{R}_{n}^{\dagger },\,\hat{S}_{K}]$
by multiplying each side of equation (\ref{Ap1.eq}) by $U_{n}(\mathbf{R}%
)\,\,V_{K}^{\beta \ast }(\mathbf{R}^{/})$, integrating $\mathbf{R,R}^{/}$
over the cavity and external regions respectively, then summing over $\beta $%
. The treatment then involves the same steps as in the case of $[\hat{Q}%
_{n},\,\hat{P}_{K}^{\dagger }]$ except for the replacement of $V_{n}^{\ast }$
by $U_{n}$ and $U_{K}^{\beta }$ by $V_{K}^{\beta \ast }$ together with the
presence of a negative sign resulting from the original equation (\ref
{Ap1.eq}) itself. This leads to: 
\begin{equation}
\lbrack \hat{R}_{n}^{\dagger },\,\hat{S}_{K}]=-\,(\mathbf{\,}\hbar
\,/\,k_{n})\,\int_{S}d^{2}\mathbf{s}\,\,\mathbf{U}_{n}(\mathbf{s,}%
z_{b})\cdot \mathbf{V}_{K}^{\ast }(\mathbf{s},z_{b}),
\end{equation}
so after taking the adjoint of both sides of this result we have finally: 
\begin{equation}
\lbrack \hat{R}_{n},\,\hat{S}_{K}^{\dagger }]=(\mathbf{\,}\hbar
\,/\,k_{n})\,\int_{S}d^{2}\mathbf{s}\,\,\mathbf{U}_{n}^{\ast }(\mathbf{s,}%
z_{b})\cdot \mathbf{V}_{K}(\mathbf{s},z_{b}).  \label{Ap12.eq}
\end{equation}

\subsubsection{\textit{Cases of }$\,[\hat{Q}_{n},\,\hat{S}_{K}^{\dagger }]$%
\textit{\ and }$[\hat{R}_{n},\,\hat{P}_{K}^{\dagger }]$}

To obtain the commutation rules for $\,[\hat{Q}_{n},\,\hat{S}_{K}^{\dagger
}] $ and $[\hat{R}_{n},\,\hat{P}_{K}^{\dagger }]$ we use equations (\ref
{RQ.eq}) to write $\hat{Q}_{n}$ as a linear combination of the $\hat{R}_{m}$
and $\hat{R}_{n}$ as a linear combination of the $\hat{Q}_{m}$. This
involves the transformation matrices $\mathbf{D}_{nm}$ and $\mathbf{C}_{nm}$
respectively. We then use the previously derived results in equations (\ref
{Ap12.eq}, \ref{Ap11.eq}) for the commutators $[\hat{R}_{m},\,\hat{S}%
_{K}^{\dagger }]$ and $[\hat{Q}_{m},\,\hat{P}_{K}^{\dagger }]$. The $(%
\mathbf{\,}\hbar \,/\,k_{m})$ terms in the sums over $m$ can be replaced by $%
(\mathbf{\,}\hbar \,/\,k_{n})$ since the transformation matrices are zero
unless $k_{n}=k_{m}$. We then use equations (\ref{UV.eq}) and the
Hermiteancy properties of $\mathbf{D}_{nm}$ and $\mathbf{C}_{nm}$ to perform
the sum over $m$, and this leads to the NHM\ functions $\mathbf{U}_{m}^{\ast
}$ and $\mathbf{V}_{m}^{\ast }$ being replaced by $\mathbf{V}_{n}^{\ast }$
and $\mathbf{U}_{n}^{\ast }$ respectively. This gives the results: 
\begin{eqnarray}
\lbrack \hat{Q}_{n},\,\hat{S}_{K}^{\dagger }] &=&(\mathbf{\,}\hbar
\,/\,k_{n})\,\int_{S}d^{2}\mathbf{s}\,\,\mathbf{V}_{n}^{\ast }(\mathbf{s,}%
z_{b})\cdot \mathbf{V}_{K}(\mathbf{s},z_{b})  \label{Ap13.eq} \\
\lbrack \hat{R}_{n},\,\hat{P}_{K}^{\dagger }] &=&(\mathbf{\,}\hbar
\,/\,k_{n})\,\int_{S}d^{2}\mathbf{s}\,\,\mathbf{U}_{n}^{\ast }(\mathbf{s,}%
z_{b})\cdot \mathbf{U}_{K}(\mathbf{s},z_{b}).  \label{Ap14.eq}
\end{eqnarray}

\subsubsection{\textit{Cases of }$\,[\hat{Q}_{n},\,\hat{P}_{K}]$\textit{\ , }%
$[\hat{R}_{n},\,\hat{S}_{K}]\,,[\hat{Q}_{n},\,\hat{S}_{K}]$\textit{, }$[\hat{%
R}_{n},\,\hat{P}_{K}]$\textit{\ and their adjoints}}

These commutators involve either two or zero adjoints and we conclude they
may be taken as equal to zero. Starting from equation (\ref{Ap1.eq}) we
could multiply each side by $V_{n}^{\ast }(\mathbf{R})\,\,V_{K}^{\beta \ast
}(\mathbf{R}^{/})$, integrate $\mathbf{R,R}^{/}$ over the cavity and
external regions respectively, then sum over $\beta $ and follow similar
steps as for $[\hat{Q}_{n},\,\hat{P}_{K}^{\dagger }]$ to obtain a surface
integral expression for $\,[\hat{Q}_{n},\,\hat{S}_{K}]$. The surface
integral would involve the scalar product $\mathbf{V}_{n}^{\ast }(\mathbf{s,}%
z_{b})\,\cdot \mathbf{V}_{K}^{\ast }(\mathbf{s,}z_{b})$ with both factors
conjugated. The cavity-external boundary is never really planar, so the
exponential factors associated with $\mathbf{V}_{n}^{\ast }(\mathbf{s,}%
z_{b})\,\cdot \mathbf{V}_{K}^{\ast }(\mathbf{s,}z_{b})$ would average out to
zero. On this basis we neglect commutators such as $[\hat{Q}_{n},\,\hat{S}%
_{K}]$, $[\hat{R}_{n},\,\hat{S}_{K}]\,$, $[\hat{Q}_{n},\,\hat{S}_{K}]$, $[%
\hat{R}_{n},\,\hat{P}_{K}]$ and their adjoints. Alternatively, using
equations (\ref{QRq.eq}) for the cavity NHM generalised coordinates and
those analogous to equation (\ref{PiTp.eq}) for the external NHM generalised
momenta, we see that such commutators would be linear combinations of true
mode commutators $[\hat{q}_{k},\hat{p}_{l}{}]$, $[\hat{q}_{k}{}^{\dagger },%
\hat{p}_{l}^{\dagger }]$ that are all zero.

\subsection{\textit{Cases of cavity NHM momenta and external NHM coordinates}%
}

The commutation rules between cavity NHM momenta and external region
coordinates are found by substituting for $\mathbf{\hat{D}(R)}$ and $\mathbf{%
\hat{B}(\mathbf{R}^{/})}$ from equations (\ref{ApDCav.eq}, \ref{ApBExt.eq})
into equation (\ref{ApCRDB.eq}), which gives:

\begin{eqnarray}
i\hbar \,\sum_{\gamma }\varepsilon _{\alpha \beta \gamma }\,\partial
/\partial X_{\gamma }\,\delta (\mathbf{R-R}^{/}) &=&\sum_{m(\alpha
)}\sum_{L}\{\,U_{m}(\mathbf{R})\,\{\mathbf{\nabla \times U}_{L}(\mathbf{R}%
^{/})\}^{\beta }\,[\hat{S}_{m},\,\hat{Q}_{L}]\,  \notag \\
&&\qquad \qquad +U_{m}(\mathbf{R})\,\{\mathbf{\nabla \times V}_{L}(\mathbf{R}%
^{/})\}^{\beta \ast }\,[\hat{S}_{m},\,\hat{R}_{L}^{\dagger }]  \notag \\
&&\qquad \qquad +V_{m}^{\ast }(\mathbf{R})\,\{\mathbf{\nabla \times U}_{L}(%
\mathbf{R}^{/})\}^{\beta }\,[\hat{P}_{m}^{\dagger },\,\hat{Q}_{L}]\,  \notag
\\
&&\qquad \qquad +V_{m}^{\ast }(\mathbf{R})\,\{\mathbf{\nabla \times V}_{L}(%
\mathbf{R}^{/})\}^{\beta \ast }\,[\hat{P}_{m}^{\dagger },\,\hat{R}%
_{L}^{\dagger }]\mathbf{\,}\}.  \notag \\
&&  \label{Ap15.eq}
\end{eqnarray}
On the right side the sum over $m$ only involves terms with $\mathbf{\hat{%
\alpha}}_{m}=\mathbf{\hat{\alpha}}$.

\subsubsection{\textit{Case of }$\,[\hat{S}_{n},\,\hat{R}_{K}^{\dagger }]$}

To obtain the commutation rule for $[\hat{S}_{n},\,\hat{R}_{K}^{\dagger }]\,$
we multiply each side of equation (\ref{Ap15.eq}) by $V_{n}^{\ast }(\mathbf{R%
})\,\,\{\mathbf{\nabla \times U}_{K}(\mathbf{R}^{/})\}^{\beta }$, integrate $%
\mathbf{R,R}^{/}$ over the cavity and external regions respectively, then
sum over $\beta $. Using the biorthogonality results in equation (\ref
{Biorth.eq}) and the results in equations (\ref{Omega.eq}, \ref{Omega2.eq})
for the magnetic energy term and (as usual) neglecting integrals where the
complex conjugation occurs either zero or two times we find that: 
\begin{eqnarray}
k_{K}^{2}\,[\hat{S}_{n},\,\hat{R}_{K}^{\dagger }] &=&\mathbf{\,}i\hbar
\,\sum_{\gamma }\varepsilon _{\alpha \beta \gamma }\,\int_{C}d^{3}\mathbf{R\,%
}\int_{E}d^{3}\mathbf{R}^{/}\,V_{n}^{\ast }(\mathbf{R})\,\{\mathbf{\nabla
\times U}_{K}(\mathbf{R}^{/})\}^{\beta }\,  \notag \\
&&\qquad \qquad \qquad \qquad \qquad \qquad .\,\partial /\partial X_{\gamma
}\,\delta (\mathbf{R-R}^{/})  \notag \\
&&  \label{Ap16.eq}
\end{eqnarray}
where $\mathbf{\alpha }_{n}=\mathbf{\hat{\alpha}}$. Since $\mathbf{\alpha }%
_{n}=\mathbf{\hat{x}}$ or $\mathbf{\alpha }_{n}=\mathbf{\hat{y}}$ are the
only two possibilities we get the two equations: 
\begin{eqnarray}
k_{K}^{2}\,[\hat{S}_{n},\,\hat{R}_{K}^{\dagger }] &=&\mathbf{\,}i\hbar
\,\int_{C}d^{3}\mathbf{R\,}\int_{E}d^{3}\mathbf{R}^{/}\,V_{n}^{\ast }(%
\mathbf{R})\,\,\{\mathbf{\nabla \times U}_{K}(\mathbf{R}^{/})\}^{y}\,%
\partial /\partial z\,\delta (\mathbf{R-R}^{/})  \notag \\
&&-\mathbf{\,}i\hbar \,\int_{C}d^{3}\mathbf{R\,}\int_{E}d^{3}\mathbf{R}%
^{/}\,V_{n}^{\ast }(\mathbf{R})\,\{\mathbf{\nabla \times U}_{K}(\mathbf{R}%
^{/})\}^{z}\,\,\partial /\partial y\,\delta (\mathbf{R-R}^{/})  \notag \\
&&  \label{Ap17.eq}
\end{eqnarray}
for $\mathbf{\alpha }_{n}=\mathbf{\hat{x}}$ and 
\begin{eqnarray}
k_{K}^{2}\,[\hat{S}_{n},\,\hat{R}_{K}^{\dagger }] &=&-\,\mathbf{\,}i\hbar
\,\int_{C}d^{3}\mathbf{R\,}\int_{E}d^{3}\mathbf{R}^{/}\,V_{n}^{\ast }(%
\mathbf{R})\,\,\{\mathbf{\nabla \times U}_{K}(\mathbf{R}^{/})\}^{x}\,%
\partial /\partial z\,\delta (\mathbf{R-R}^{/})  \notag \\
&&+\mathbf{\,}i\hbar \,\int_{C}d^{3}\mathbf{R\,}\int_{E}d^{3}\mathbf{R}%
^{/}\,V_{n}^{\ast }(\mathbf{R})\,\,\{\mathbf{\nabla \times U}_{K}(\mathbf{R}%
^{/})\}^{z}\,\partial /\partial x\,\delta (\mathbf{R-R}^{/})  \notag \\
&&  \label{Ap18.eq}
\end{eqnarray}
for $\mathbf{\alpha }_{n}=\mathbf{\hat{y}}$.

As previously, the integrals involving $\partial /\partial x\,$\ or $%
\partial /\partial y$ give zero, so only the integrals involving $\partial
/\partial z$ remain. Following the same treatment for the latter results in
expressions involving surface integrals: 
\begin{equation}
k_{K}^{2}\,[\hat{S}_{n},\,\hat{R}_{K}^{\dagger }]=\mathbf{\,}i\hbar
\,\int_{S}d^{2}\mathbf{s}\,\,V_{n}^{\ast }(\mathbf{s,}z_{b})\,\{\nabla
\times \mathbf{U}_{K}(\mathbf{s},z_{b})\}^{y}
\end{equation}
for $\mathbf{\alpha }_{n}=\mathbf{\hat{x}}$ and 
\begin{equation}
k_{K}^{2}\,[\hat{S}_{n},\,\hat{R}_{K}^{\dagger }]=-\mathbf{\,}i\hbar
\,\int_{S}d^{2}\mathbf{s}\,\,V_{n}^{\ast }(\mathbf{s,}z_{b})\,\{\nabla
\times \mathbf{U}_{K}(\mathbf{s},z_{b})\}^{x}
\end{equation}
or $\mathbf{\alpha }_{n}=\mathbf{\hat{y}}$. \ These results may be
summarised in a single formula: 
\begin{equation}
\lbrack \hat{S}_{n},\,\hat{R}_{K}^{\dagger }]=\mathbf{\,(}i\hbar
\,/k_{K}^{2})\,\int_{S}d^{2}\mathbf{s}\,\,\mathbf{\hat{z}\cdot V}_{n}^{\ast
}(\mathbf{s,}z_{b})\,\times \{\nabla \times \mathbf{U}_{K}(\mathbf{s}%
,z_{b})\}.  \label{Ap19.eq}
\end{equation}

\subsubsection{\textit{Case of }$\,[\hat{P}_{n},\,\hat{Q}_{K}^{\dagger }]$}

To obtain the commutation rule for $[\hat{P}_{n},\,\hat{Q}_{K}^{\dagger }]$,
we first obtain the result for $[\hat{P}_{n}^{\dagger },\,\hat{Q}_{K}]$ by
multiplying each side of equation (\ref{Ap15.eq}) by $U_{n}(\mathbf{R})\,\,\{%
\mathbf{\nabla \times V}_{K}(\mathbf{R}^{/})\}^{\beta \ast }$, integrating $%
\mathbf{R,R}^{/}$ over the cavity and external regions respectively, then
summing over $\beta $. The treatment then involves the same steps as in the
case of $[\hat{S}_{n},\,\hat{R}_{K}^{\dagger }]$ except for the replacement
of $V_{n}^{\ast }$ by $U_{n}$ and $\{\mathbf{\nabla \times U}_{K}(\mathbf{R}%
^{/})\}^{\beta }$ by $\{\mathbf{\nabla \times V}_{K}(\mathbf{R}%
^{/})\}^{\beta \ast }$. This leads to: 
\begin{equation}
\lbrack \hat{P}_{n}^{\dagger },\,\hat{Q}_{K}]=\mathbf{\,(}i\hbar
\,/k_{K}^{2})\,\int_{S}d^{2}\mathbf{s}\,\,\mathbf{\hat{z}\cdot U}_{n}(%
\mathbf{s,}z_{b})\,\times \{\nabla \times \mathbf{V}_{K}^{\ast }(\mathbf{s}%
,z_{b})\},
\end{equation}
and taking the adjoint of both sides leads to the result: 
\begin{equation}
\lbrack \hat{P}_{n},\,\hat{Q}_{K}^{\dagger }]=\mathbf{\,(}i\hbar
\,/k_{K}^{2})\,\int_{S}d^{2}\mathbf{s}\,\,\mathbf{\hat{z}\cdot U}_{n}^{\ast
}(\mathbf{s,}z_{b})\,\times \{\nabla \times \mathbf{V}_{K}(\mathbf{s}%
,z_{b})\}.  \label{Ap20.eq}
\end{equation}

\subsubsection{\textit{Cases of }$\,[\hat{P}_{n},\,\hat{R}_{K}^{\dagger }]$%
\textit{\ and }$[\hat{S}_{n},\,\hat{Q}_{K}^{\dagger }]$}

To obtain the commutation rules for $\,[\hat{P}_{n},\,\hat{R}_{K}^{\dagger
}] $ and $[\hat{S}_{n},\,\hat{Q}_{K}^{\dagger }]$ we use equations (\ref
{PiT.eq}) to write $\hat{P}_{n}$ as a linear combination of the $\hat{S}_{m}$
and $\hat{S}_{n}$ as a linear combination of the $\hat{P}_{m}$. The same
method used for determining $\,[\hat{Q}_{n},\,\hat{S}_{K}^{\dagger }]$ and $[%
\hat{R}_{n},\,\hat{P}_{K}^{\dagger }]$ is then followed and we find that: 
\begin{eqnarray}
\lbrack \hat{P}_{n},\,\hat{R}_{K}^{\dagger }] &=&\mathbf{(}i\hbar
\,/k_{K}^{2})\,\int_{S}d^{2}\mathbf{s}\,\,\mathbf{\hat{z}\cdot U}_{n}^{\ast
}(\mathbf{s,}z_{b})\,\times \{\nabla \times \mathbf{U}_{K}(\mathbf{s}%
,z_{b})\}  \label{Ap21.eq} \\
\lbrack \hat{S}_{n},\,\hat{Q}_{K}^{\dagger }] &=&\mathbf{(}i\hbar
\,/k_{K}^{2})\,\int_{S}d^{2}\mathbf{s}\,\,\mathbf{\hat{z}\cdot V}_{n}^{\ast
}(\mathbf{s,}z_{b})\,\times \{\nabla \times \mathbf{V}_{K}(\mathbf{s}%
,z_{b})\}.  \label{Ap22.eq}
\end{eqnarray}

\subsubsection{\textit{Cases of }$\,[\hat{P}_{n},\,\hat{Q}_{K}]$\textit{\ , }%
$[\hat{S}_{n},\,\hat{R}_{K}]\,,[\hat{P}_{n},\,\hat{R}_{K}]$\textit{, }$[\hat{%
S}_{n},\,\hat{Q}_{K}]$\textit{\ and their adjoints}}

These commutators involve either two or zero adjoints and we conclude they
may be taken as equal to zero. The details are similar to those for $[\hat{Q}%
_{n},\,\hat{P}_{K}]$ , $[\hat{R}_{n},\,\hat{S}_{K}]\,,[\hat{Q}_{n},\,\hat{S}%
_{K}]$, $[\hat{R}_{n},\,\hat{P}_{K}]$ and their adjoints.

\section{Figure captions}

Figure 1. Unstable optical resonator system consisting of two confocal
mirrors. The dashed line indicates the boundary between the cavity and the
external regions. The left hand mirror is assumed to be large, so that light
does not travel past it and is confined to the cavity and external regions
shown.

\noindent Figure 2. Ranges of $z$ integration for cavity and external
regions used to determine cavity NHM/external NHM commutation rules.

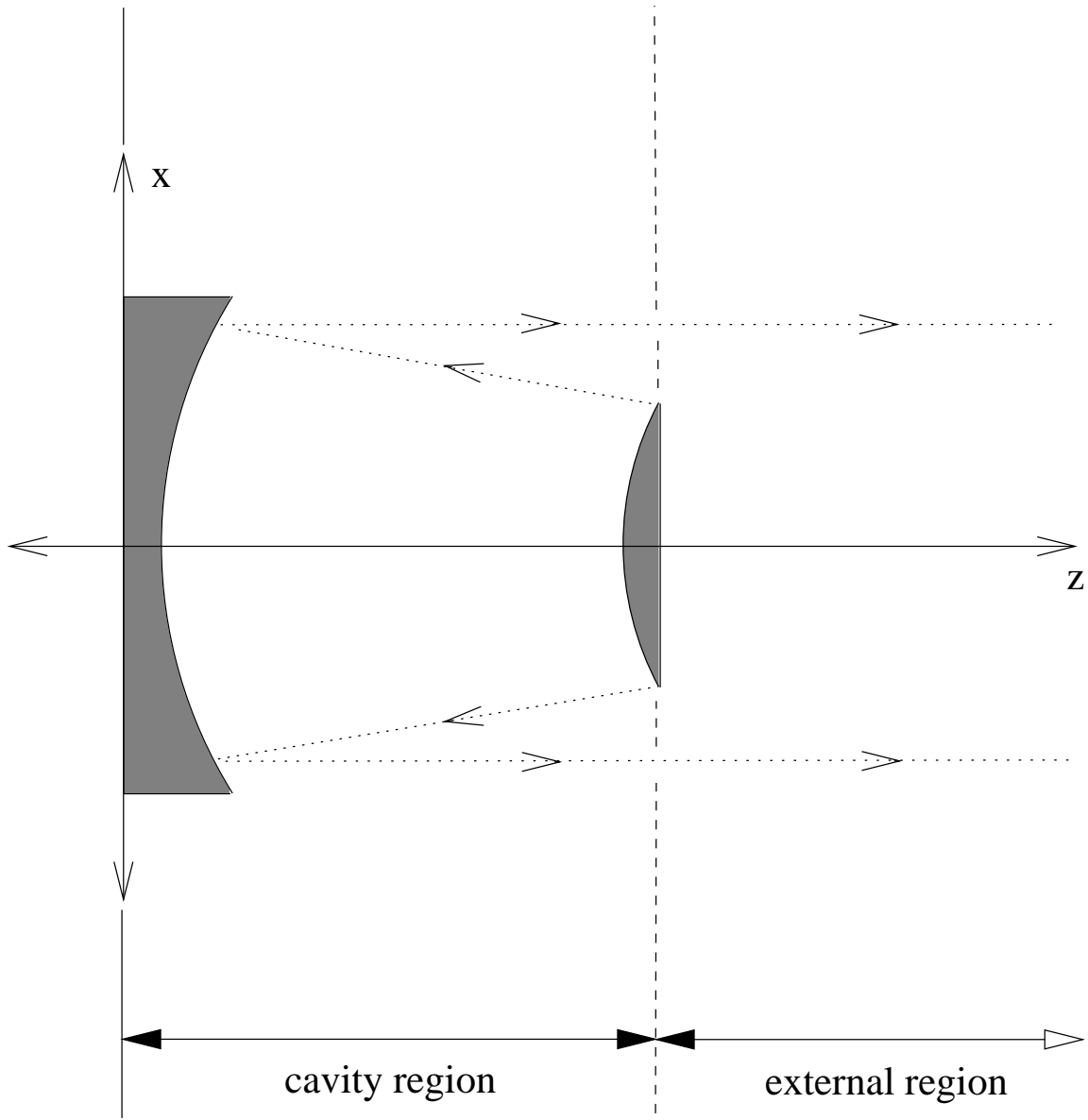
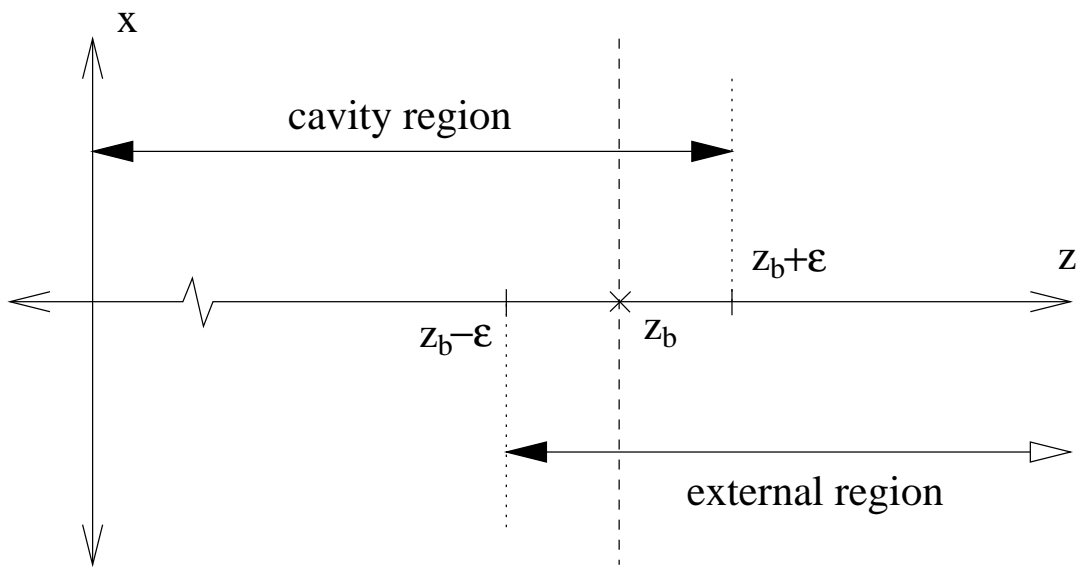
\end{document}